\documentclass[journal]{IEEEtran}
\usepackage[utf8]{inputenc}

\usepackage{setspace}
\newcommand{\graphixpath}{./Figures/}
\usepackage{epsfig}
\usepackage{amsmath}
\usepackage{graphicx}
\usepackage{graphics}
\usepackage{subcaption}
\usepackage{multirow}
\usepackage{authblk}
\usepackage{url}
\usepackage{booktabs}
\usepackage{mathtools}
\usepackage{multicol}
\newcommand{\ra}[1]{\renewcommand{\arraystretch}{#1}}
\usepackage{cite}

 \title{Trajectory Optimization of Robots with Regenerative Drive Systems: Numerical and Experimental Results}

\author{Poya~Khalaf,
        and~Hanz~Richter
\thanks{P. Khalaf is a Doctoral Student with the Department of Mechanical Engineering, Cleveland State University, e-mail: p.khalaf@csuohio.edu.}
\thanks{H. Richter is a Professor with the Department of Mechanical Engineering, Cleveland State University, e-mail: h.richter@csuohio.edu.}}

\begin{document}

\maketitle

\begin{abstract}
We investigate energy-optimal control of robots with ultracapacitor based regenerative drive systems. Based on a previously introduced framework, a fairly generic model is considered for the robot and the drive system. An optimal control problem is formulated to find point-to point trajectories maximizing the amount of energy regenerated and stored in the capacitor. The optimization problem, its numerical solution and an experimental evaluation are demonstrated using a PUMA 560 manipulator. A comprehensive experimental setup was prepared to evaluate power flows and energy regeneration. Tracking of optimal trajectories was enforced on the robot using a standard robust passivity based control approach. Experimental results show that when following optimal trajectories, a reduction of about $13\%$ in energy consumption can be achieved for the conditions of the study.
\end{abstract}

\section{Introduction}
\IEEEPARstart{E}nergy regeneration technologies have gained much attention due to their potential to reduce the energy consumption of modern engineering systems. Lower energy consumption allows devices to work for longer periods of time with lower operational costs. These factors are crucial in
the design of systems such as electric and hybrid vehicles~\cite{lukic2006power}, powered prostheses and exoskeletons~\cite{hitt2009robotic}, autonomous spacecraft~\cite{shimizu2013super}, and others. The concept of energy regeneration is understood here to be the process of recovering energy that would be otherwise dissipated and redistributing or storing it for later use. The framework proposed in~\cite{richter2015framework} allows energy optimization and motion controller designs to be conducted separately, by introducing a virtual controller and capacitor voltage feedback.

This work is motivated by the application of regenerative technologies in robotic systems. Incorporating regenerative design features in robotic systems is justified when a significant potential for energy recovery exists. Examples include fast-moving, multi-joint industrial and mobile robots, powered prostheses and powered exoskeletons. Excess energy can be stored from the robot joints when decelerating and reused when the robot joints are accelerating thus, reducing the overall energy consumption. For an industrial manufacturing line with many robotic systems, this can lead to a significant reduction in electric power costs. For powered prostheses and exoskeletons, energy regeneration can increase operating time therefore, making them more practical for daily use.

In addition, robots with regenerative drive systems offer unique opportunities for joint-to-joint mechanical energy redistribution by electrical means. Strictly speaking, energy transfer among joints may naturally occur in robotic manipulators via inertial coupling. However, this kind of indirect energy transfer is governed by the structure, mass properties, and joint trajectories of the robot. In many cases, these factors are predefined and the joint-to-joint energy flow cannot be managed or controlled. For example, the structure of a Cartesian robot prevents any energy flow from one joint to another.
Bidirectional power (4-quadrant) drive electronics offer the opportunity to configure pathways for joint-to-joint energy transfer and management. Excess energy regenerated from a joint decelerating can be conveyed to another joint that is accelerating and demanding energy. In a regenerative Cartesian robot, this allows for direct energy transfer between joints. Such capabilities can lead to significant reduction in the energy consumption of the overall robot. 

We consider regenerative drive systems that use capacitive means for storing energy. The development of electrochemical double layer capacitors, so-called ultracapacitors or supercapacitors, in the past decade have enabled efficient means of storing and reusing energy~\cite{conway2013electrochemical}. Unlike batteries, ultracapacitors can be charged and discharged at high rates without damaging them and have considerably high power densities\cite{khaligh2010battery}. Being lightweight, inexpensive and durable are other properties of ultracapacitors. Because of these properties, ultracapacitors are being used in many applications involving energy regeneration~\cite{shimizu2013super,khoshnoud2015energy,vinot2013optimal,song2014energy,rufer2002supercapacitor,zhang2016high}. 

The research literature is replete with papers discussing energy regeneration and the use of ultracapacitors in systems such as road vehicles~\cite{vinot2013optimal,song2014energy,lukic2006power,khaligh2010battery},  industrial electric motor drive systems~\cite{grbovic2011modeling,rufer2002supercapacitor,grbovic2011ultracapacitor}, vibration control and shock absorber systems~\cite{khoshnoud2015energy,kammer2016enhancing,asai2016nonlinear,zhang2016high}, aerospace applications~\cite{shimizu2013super} and so on. However, research regarding use of these technologies in robotic systems is scarce. Here we offer a brief review and refer readers to the recent survey~\cite{carabin17} for a more comprehensive study of the literature.


Izumi et al.~\cite{izumi1995optimal} considered a DC servo system capable of regenerating excess energy into a conventional capacitor. They formulated and solved a point-to-point trajectory optimization problem for this servo system by minimizing the dissipated energy. Experimental results showed storage of excess energy in the capacitor while the motor was decelerating . In a later work Izumi et al.~\cite{izumi2000energy} considered a two-link vertically articulated  manipulator with energy regeneration. A point-to point optimal trajectory problem minimizing dissipated energy was solved for this robot. Simulation results showed that the optimal trajectory reduces energy consumption compared to the conventional non-optimized trajectory. While conventional capacitors were used, the authors pointed out the need for larger capacitances.

Fujimoto~\cite{fujimoto2004minimum} found energy minimizing trajectories for bipedal running. The problem was formulated as an optimal control problem and solved numerically for a five link planar biped robot. The analysis took into account the possibility of energy regeneration. The optimal knee trajectory showed regions of positive and negative power. Based on the optimization results it was concluded that the use of energy regeneration mechanisms, such as elastic actuators or back-drivable actuators combined with bidirectional power converters, can be used to reduce the over all energy consumption.

Hansen et al.~\cite{hansen2012enhanced} considered finding trajectories for a KUKA robot that minimize the amount of external electrical energy supplied to the motor drivers. The motor drivers are coupled together through a common DC bus, allowing power to flow from one joint of the robot to another. However their work does not include a capacitor to store excess energy. Thus energy regenerated by the robot joints is wasted unless at the same time other joints utilize the regenerated energy. The authors point out the use of a storage capacitor moving forward. Joint trajectories are described by B-splines and are optimized using a gradient based optimization method. Experimental results showed a $10\%$ decrease in total energy consumption for the robot.  

In the robotic lower limb prostheses field, a group from the Massachussetts Institute of Technology pioneered a regenerative transfemoral prosthesis in the 1980’s that used conventional capacitors to store regenerated energy~\cite{hunter1981design,seth1987energy,tabor1988real}. They aimed to design the system so that no external power would be required for operation. The power required for the prosthesis would be regenerated during the passive portions of gait. Results indicated the need for larger capacitances, which were not available at the time. More recent work investigate the use of elastic elements to store the regenerated energy~\cite{hitt2009robotic,hitt2010active,hitt2007sparky,holgate2008sparky,everarts2012variable}. Tucker et al.~\cite{tucker2010mechanical} developed an analytical model of a regenerative powered transferral prosthesis. Energy is regenerated by controlling the actuator damping during its passive regions of operation. A regeneration manifold is found that limits the actuator damping which can be achieved, while regenerating energy. 

Richter~\cite{richter2015framework} proposed a unifying framework for modeling and control of robots with regenerative drive systems. It enables a systematic treatment of robot motion control with explicit consideration of energy regeneration.  It is capable of capturing various regenerative actuators in various domains (e.g., electromechanical, hydraulic, etc.). Based on this framework, several papers have focused on the use of evolutionary algorithms and other numerical methods to find optimal system parameters that optimize a combination of motion tracking and energy enhancing objectives~\cite{warner2015optimal,warner2016IEEE,rarick2014optimal,rohani2017optimal}. In addition, the authors have investigated analytical solutions to the general parameter optimization for robotic systems~\cite{khalaf2018global,khalaf2016parametric}. Global closed form solutions are found for the robot parameters (e.g. link lengths) and drive mechanism parameters (e.g. gear ratios) that maximize energy regeneration given a predefined fixed trajectory for the robot. While most efforts have focused on theory and simulations, 
experimental evaluations of the effectiveness of energy regeneration are very scarce in the robotics literature. 

In this paper a generic robotic model with ultracapacitor based regenerative drive systems is developed using the framework of~\cite{richter2015framework}. The problem of finding optimal point-to-point trajectories that maximize regenerative energy storage is formulated as an optimal control problem and demonstrated with a PUMA 560 robot. 
The direct collocation method~\cite{von2013numerical} is used to find the optimal trajectories. The optimal trajectories are then implemented on a PUMA 560 robot using a semi-active virtual control (SVC) strategy~\cite{richter2014semiactive,richter2015framework} and a standard robust passivity based controller~\cite{SHV}. An experimental setup is prepared to evaluate the effectiveness of energy regeneration by measuring power flows at key locations.

Section~\ref{sec:modeling} discusses modeling of the robotic system and the regenerative drive mechanism, Section~\ref{sec:frmoptprb} formulates the point-to-point trajectory optimal control problem, Section~\ref{sec:optres} discuses the optimization results for the PUMA 560 robot, Section~\ref{sec:expeval} explains the experimental procedure for evaluating the optimized trajectories and discusses the experimental results, and Section~\ref{sec:conclusion} presents some concluding remarks and possible paths for future work.

\section{Modeling}\label{sec:modeling}
We consider general serial robots modeled with the dynamic equations: 
\begin{equation}
D^\circ(q) \ddot{q}+C(q,\dot{q}) \dot{q} +\mathcal{R}^\circ(q,\dot{q})+g(q)+\mathcal{T}=\tau
\label{manipulator dynamic}
\end{equation}
where $q$ is the $n\times 1$ vector of joint coordinates, $D(q)$ is the inertia matrix, $C(q,\dot{q})$ is a matrix accounting for Coriolis and centrifugal effects, $\mathcal{R}(q,\dot{q})$ is a general nonlinear damping term, $\mathcal{T}$ is the vector of external forces and moments reflected to the manipulator joints, $g(q)$ is the gravity vector and $\tau$ is the vector of joint forces and moments applied by a set of actuators.

\begin{figure}[!t]
\centering
\includegraphics[width=2.5in,trim=0mm 0mm 0mm 30mm,clip=true]{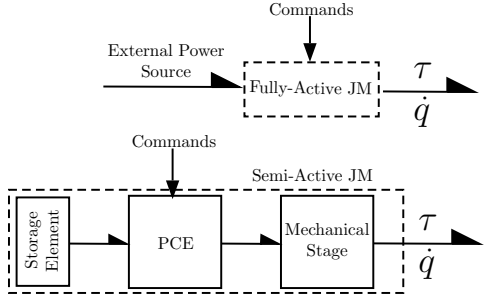}
\caption{Schematic of fully-active and semi-active joint mechanisms. Fully-active joint mechanisms uses external power for actuation while semi-active joint mechanisms use an energy storing element and only exchange mechanical power with the robot.}
\label{JMdef} 
\end{figure}

In this context, the robot actuators are either conventional (termed \textit{fully-active}) or regenerative (termed \textit{semi-active}). A fully-active actuator is conventional in the sense that it exchanges mechanical power with the robot and draws electric power from an external source (similar to typical electric drives). On the other hand, semi-active actuators have self-contained energy storage. They are passive systems and only exchange mechanical power with the robot~\cite{richter2015framework}. Figure~\ref{JMdef} depicts the concepts of fully-active and semi-active actuators. Semi-active actuators are composed of a storage device to provide energy to the robot and possibly store excess energy, a power conversion element (PCE) to regulate power and to convert power between different domains, and a mechanical stage to interface with the robot.

In a general setting, a subset of manipulator joints are assumed to be semi-active, while the remaining joints are fully-active. Also, for simplicity, the terms \textit{actuators} and \textit{joints} are used interchangeably.

\begin{figure}[!t]
\centering
\includegraphics[width=2.5in]{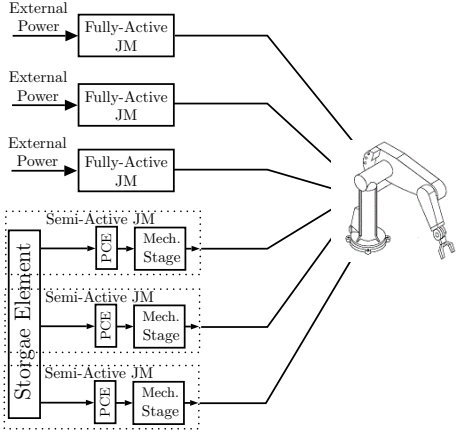}
\caption{Star configuration for semi-active joints. All the semi-active joints are connected to a common storage element. This allows for  energy transfer from one semi-active joint to another.}
\label{satrconfig1} 
\end{figure}

Depending on the arrangement of the storage elements for semi-active joints, different configurations are possible~\cite{khalaf2018global}. We consider here the \textit{star configuration} which consists of a single storage element connected in parallel to all the semi-active joints, Fig.~\ref{satrconfig1}. The star configuration provides a way to transfer power from one joint to another joint requiring energy using the common storage element as an energy reservoir.   
Other configurations for semi-active joints are possible, for instance a distributed arrangement where each semiactive joint uses a storage unit~\cite{khalaf2018global}. 

\subsection{Semi-active Actuator Modeling}
\begin{figure*}[!t]
\centering
\includegraphics[width=5in]{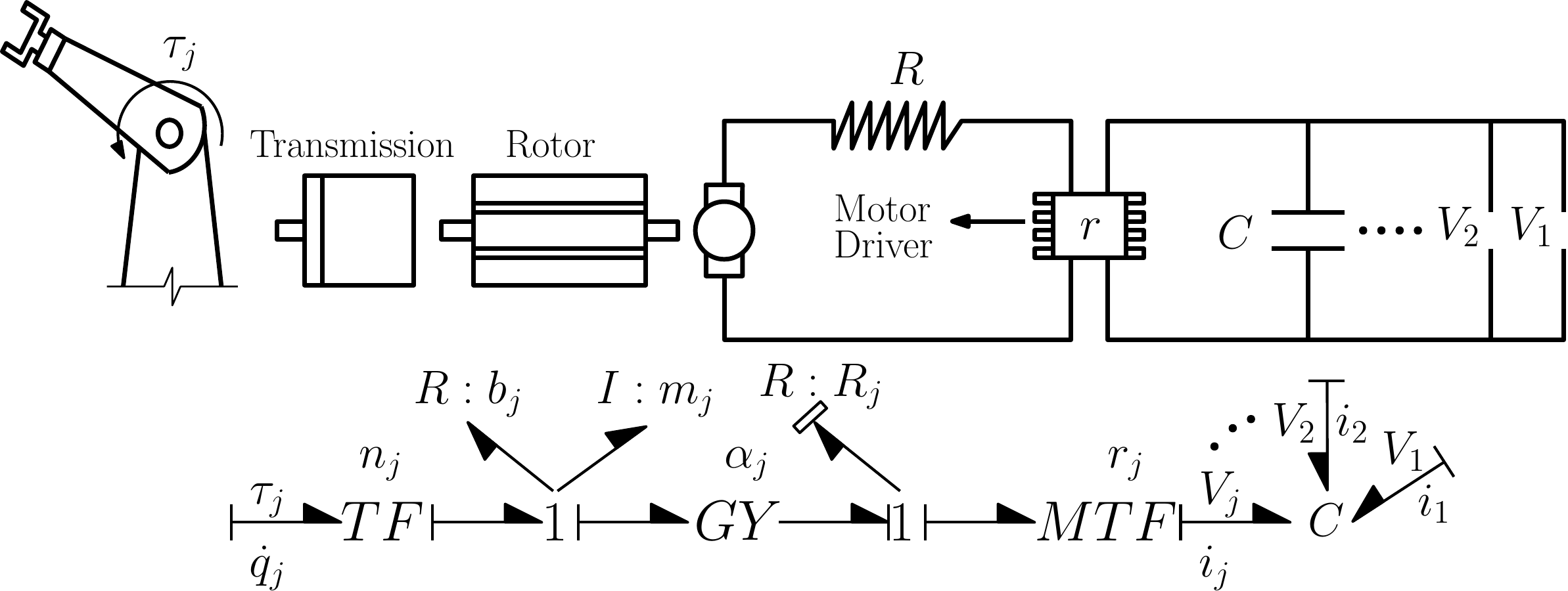}
\caption{Bond graph of electro-mechanical semi-active joint mechanism in the star configurations.}
\label{semiJMbond} 
\end{figure*}

Bond graphs~\cite{karnopp2012system} are used to facilitate the representation and equation derivation. We consider electro-mechanical semi-active actuators with an ultracapacitor as the storing element and a DC motor/generator as the PCE. The bond graph model however, can capture a wide variety of actuators in different domains (hydraulic, pneumatic, etc.). Figure~\ref{semiJMbond} shows the bond graph model of the semi-active joint in the star configuration.

Each link of the robot with a semi-active joint is connected to a transmission where $n_j$ is the velocity ratio, $m_j$ is the inertia, and $b_j$ is the viscous damping coefficient. The output of the transmission is connected to a DC motor/generator with torque constant $\alpha_j$ (which equals the back-emf constant). The inertial and frictional effects of the motor/generator are assumed to have been reflected to the link side, and already included in $m_j$ and $b_j$. Power transferred to the electrical side of the motor/generator is distributed as resistive losses and as stored energy in the ultracapacitor $C$. An ideal four quadrant motor driver is used used to control the amount and direction of voltage applied to the DC motor where $r_j$ is the converter voltage ratio (motor voltage over capacitor voltage). Since the motor driver does not boost the capacitor voltage, $r_j$ is assumed to be constrained to $[-1,1]$. A value $r_j<0$ is used to apply reverse voltage to the  DC motor terminals even though the capacitor voltage is always positive.

\subsection{Augmented Model}
The interfacing torque or force, $\tau_j$, for the $j$-th semi-active joint is derived from the bond graph model in Fig.~\ref{semiJMbond}
\begin{equation}
\tau_j=-m_jn_j^2\ddot{q}_j-(b_jn_j^2+\frac{a_j^2}{R_j})\dot{q}_j+\frac{a_jr_j}{R_j}V_{cap} 
\label{tau_aug}
\end{equation}
where $V_{cap}$ is the capacitor voltage, and $a_j=\alpha_j n_j$. Replacing $\tau_j$ from Eq.~\eqref{tau_aug} into equation Eq.~\eqref{manipulator dynamic} and absorbing the terms containing $\dot{q}$ and $\ddot{q}$ into the right-hand side, the augmented model is obtained
\begin{equation}
D(q)\ddot{q}+C(q,\dot{q}) \dot{q} +\mathcal{R}(q,\dot{q})+g+\mathcal{T}=u
\label{augmodel}
\end{equation}
where $D$ and $\mathcal{R}$ are
\begin{equation}
\begin{array}{l r}
D_{ij} = D^\circ_{ij} & i\neq j\\
\mathcal{R}_j = \mathcal{R}^\circ_j & j\not\in \{1,\dots,e\}\\
D_{jj} = D^\circ_{jj} + m_jn^2_j\ddot{q}_j & j\in \{1,\dots,e\} \\
\mathcal{R}_j = \mathcal{R}^\circ_j + (b_jn^2_j +\frac{a^2_j}{R_j})\dot{q}_j & j\in \{1,\dots,e\}\\
\end{array}
\end{equation}
and
\begin{equation}
u=
\begin{dcases*}
u_j                       & Joint $j$ is fully-active\\
\frac{a_jr_j}{R_j}V_{cap} & Joint $j$ is semi-active
\end{dcases*} 
\end{equation}
Fully-active joints are directly controlled with $u_j$, which is typically an analog input voltage to a torque-mode servo amplifier.  For the semi-active joints, only the voltage ratio $r_j$ is available as a control variable. Control is achieved with the semi-active virtual control method summarized next.

\subsection{Semi-active Virtual Control Strategy}
To control a robot with fully-active and semi-active joints, a \emph{virtual control law} ($\tau^d$) is first designed for $u$ in the augmented model (Eq.~\eqref{augmodel}). For fully-active joints, this law is enforced directly, using externally-powered servo drives. For semi-active joints, the control input $r_j$ is adjusted such that the following \emph{virtual matching} relation holds:
\begin{equation}
\frac{a_jr_j}{R_j}V_{cap}=\tau_j^d
\label{taud}
\end{equation}
The virtual control ($\tau^d$) can be any feedback law compatible with the desired motion control objectives for the augmented model. If virtual matching holds exactly Eq.~\eqref{taud} at all times, any properties that apply to the virtual design such as stability, tracking performance, robustness, etc. will be propagated to the actual system~\cite{richter2015framework}. The modulation law for exact virtual matching is simply obtained by solving for $r_j$ from Eq.~\eqref{taud}. Virtual matching is always possible as long as there is a positive voltage in the capacitor, and it will hold exactly whenever $a_j/R_j$ is precisely known and the calculated $r_j$ is within $[-1,1]$. Also, note that the virtual control law (Eq.~\eqref{taud}) and the augmented model (Eq.~\eqref{augmodel}) were derived \emph{without the need to model the ultracapacitor}. Ultracapacitor models are in general complex and nonlinear and do not cover all the aspects of the ultracapacitor's performance \cite{grbovic2013ultra,buller2001modeling,chiang2013dynamic,bertrand2010fractional}. Placing the capacitor voltage in feedback of the virtual control law, allows the analysis and control of ultracapacitor based dynamic systems without modeling complexities associated with    ultracapacitors. Furthermore, as with any system with finite on-board power storage, operation must be stopped once charge (indicated by $V_{cap}$) drops below an acceptable threshold and the system recharged. It is important to note that self-sustained operation or even charge buildup can occur, depending on system parameters and trajectories~\cite{khalaf2018global,khalaf2016parametric,richter2015framework,richter2014semiactive}.

\subsection{Regenerated Energy}
The energy provided to or extracted from the capacitor by the $j$-th semi-active joint ($\Delta E_j$) can be derived from the bond graph representation of Fig.~{\ref{semiJMbond}},
\begin{equation}
	\Delta E_j= \int_{t_1}^{t_2}v_{_j}i_{j}\,dt
	\label{deltaE1}
\end{equation}
where in the star configuration, $v_j$ is equal to the capacitor voltage $V_{cap}$, and 
\begin{equation}
	i_{j}=\frac{r_j}{R_j}\left(a_j\dot{q}_j-r_jv_{j}\right)	
\end{equation}
Replacing for $i_j$ and $v_j$ in Eq.~\eqref{deltaE1},
\begin{equation}
\Delta E_j=\int_{t_1}^{t_2}\frac{r_j}{R_j}\left(a_jV_{cap}\dot{q}_j-r_jV_{cap}^2\right)\,dt
\end{equation}
Using Eq.~\eqref{taud}, $\Delta E_j$ can be written in terms of $\tau^d$
\begin{equation}
	\Delta E_j=\int_{t_1}^{t_2}\left(\tau_j^d\dot{q}_j-\frac{R_j}{a^2_j}\left(\tau_j^d\right)^2\right)\,dt	
\end{equation}
By adding the energies provided to the capacitor by all the semi-active joints, the total energy is found to be 
\begin{equation}
	\Delta E=\int_{t_1}^{t_2}\sum_{j=1 }^{e}\left(\tau_j^d\dot{q}_j-\frac{R_j}{a^2_j}\left(\tau_j^d\right)^2\right)\,dt	
	\label{deltaE}
\end{equation}
A value of $\Delta E>0$ indicates energy regeneration and $\Delta E<0$ indicates energy consumption. Note again that as a result of SVC, the above derivation is independent of the ultracapacitor model and is a only a function of the control law $\tau^d$, joint velocities $\dot{q}$, and joint parameters $R$ and $a$. In other words, SVC decouples the dynamics of the robot and energy regeneration from the dynamics of the ultracapacitor. 

An external energy balance for the whole robotic system can be derived as
\begin{equation}
W_{act}=W_{ext}+\Delta E_m^T+\Sigma_m^T+\Delta E_s+\Sigma_e
\label{extenergy}
\end{equation}
where $W_{act}$ is the work done by the fully-active joints, $W_{ext}$ is the work done by the external forces and moments, $\Delta E_m^T$  and $\Sigma_m^T$ are the total mechanical energy and mechanical losses of the robot and the semi-active joints, respectively, $\Delta E_s$ is the energy stored in the capacitor and $\Sigma_e$ is the Joule losses of the semi-active joints. This equation shows that the energy stored in the capacitor is the net result of $W_{act}$, $W_{ext}$ and $\Delta E_m^T$ minus all losses. The derivation of Eq.\eqref{extenergy} is straightforward but omitted here for conciseness. In Section~\ref{sec:frmoptprb}, we formulate a trajectory optimization problem based on maximizing Equation~\eqref{deltaE}.

\section{Formulating the Optimization Problem}\label{sec:frmoptprb}
We aim to find trajectories for the robotic manipulator that maximize the amount of energy regenerated. For this purpose, the problem is formulated as an optimal control problem of finding the vector of optimal trajectories($q(t),\ \dot{q}(t),\ {\ddot{q}(t)}$) and the vector of optimal controls ($\tau^d$) that maximize Eq.~\eqref{deltaE}
\begin{equation}
 \underset{\tau^d\, q}{\max}\ \ J=\int_{t_1}^{t_2}\sum_{j=1 }^{e}\left(\tau^d_j\dot{q}_j-\frac{R_j{\tau^d_j}^2}{a_j^2}\right)\,dt
 \label{costfunc}
\end{equation}
while being subjected to the dynamic equations of the robot (Eq.~\eqref{augmodel}), bounds on the control, and  constraints for the starting and ending points of the trajectories.
\begin{subequations}
\begin{align}
&D(q) \ddot{q}+C(q,\dot{q}) \dot{q} +\mathcal{R}(q,\dot{q})+g(q)+\mathcal{T}=\tau^d & \label{consta}\\
&\frac{-a_j}{R_j}V_{cap}\leq \tau_j^d \leq \frac{a_j}{R_j} V_{cap} &\\
&q_{start}=q_i\ \ \ \  \dot{q}_{start}=\dot{q}_i&  \\
&q_{end}=q_f\ \ \ \ \  \dot{q}_{end}=\dot{q}_f&   
\end{align}
\label{const}
\end{subequations}
The bounds for the controls $\tau^d$ are obtained from the requirement $|r_j| \leq 1$ and Eq.\ref{taud}, where the available capacitor voltage 
$V_{cap}$ is assumed constant for the purposes of the optimization. Trajectories start from the initial position $q_{start}$ and initial velocity $\dot{q}_{start}$ and end at the final position $q_{end}$ with final velocity $\dot{q}_{end}$.

\begin{figure}[!t]
\centering
\begin{subfigure}[!t]{0.49\textwidth}
\centering
\includegraphics[width=0.2\textheight]{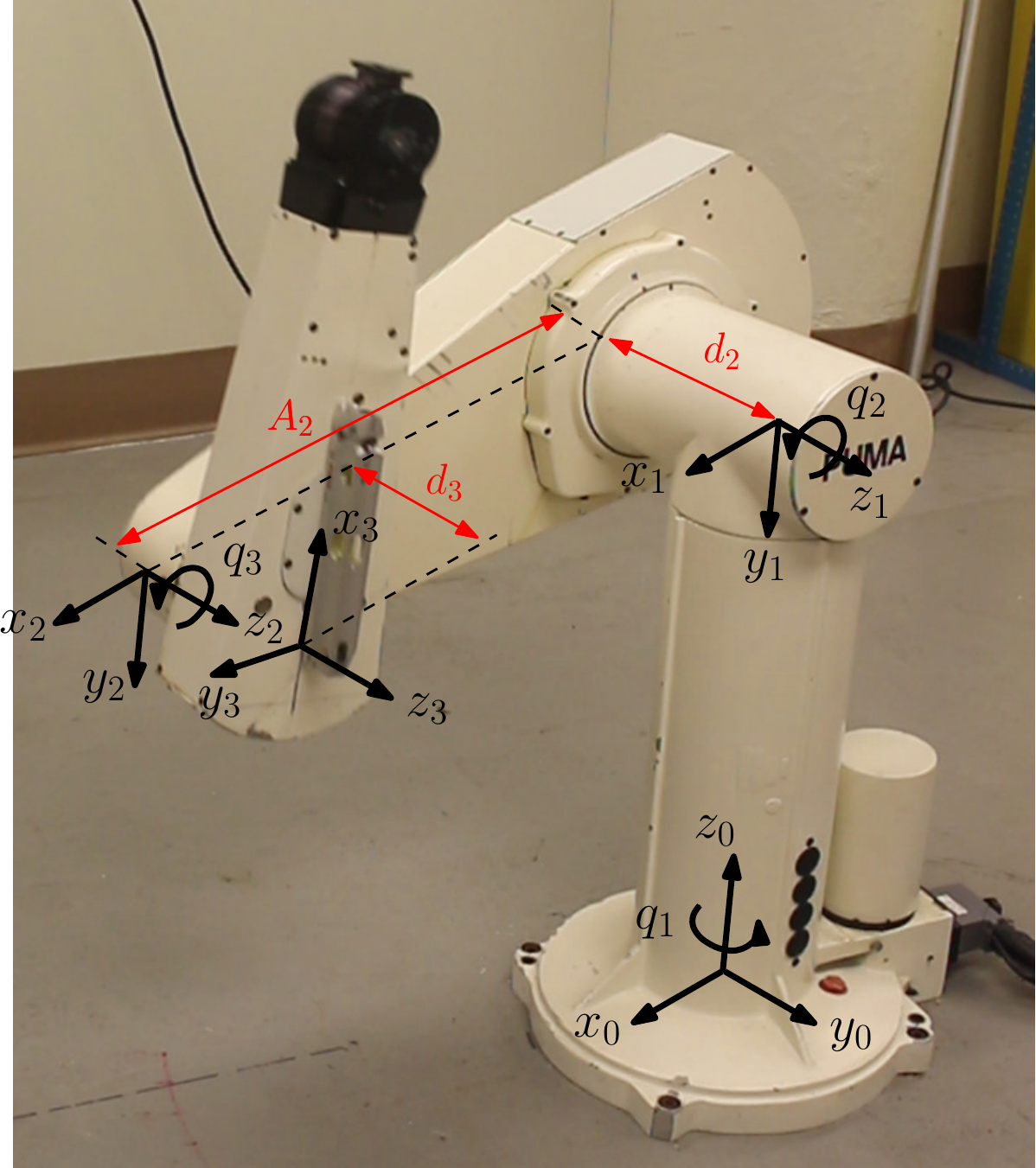}
\caption{}
\label{puma1}
\end{subfigure}
\begin{subfigure}[t]{0.49\textwidth}
\centering
\includegraphics[height=0.2\textheight]{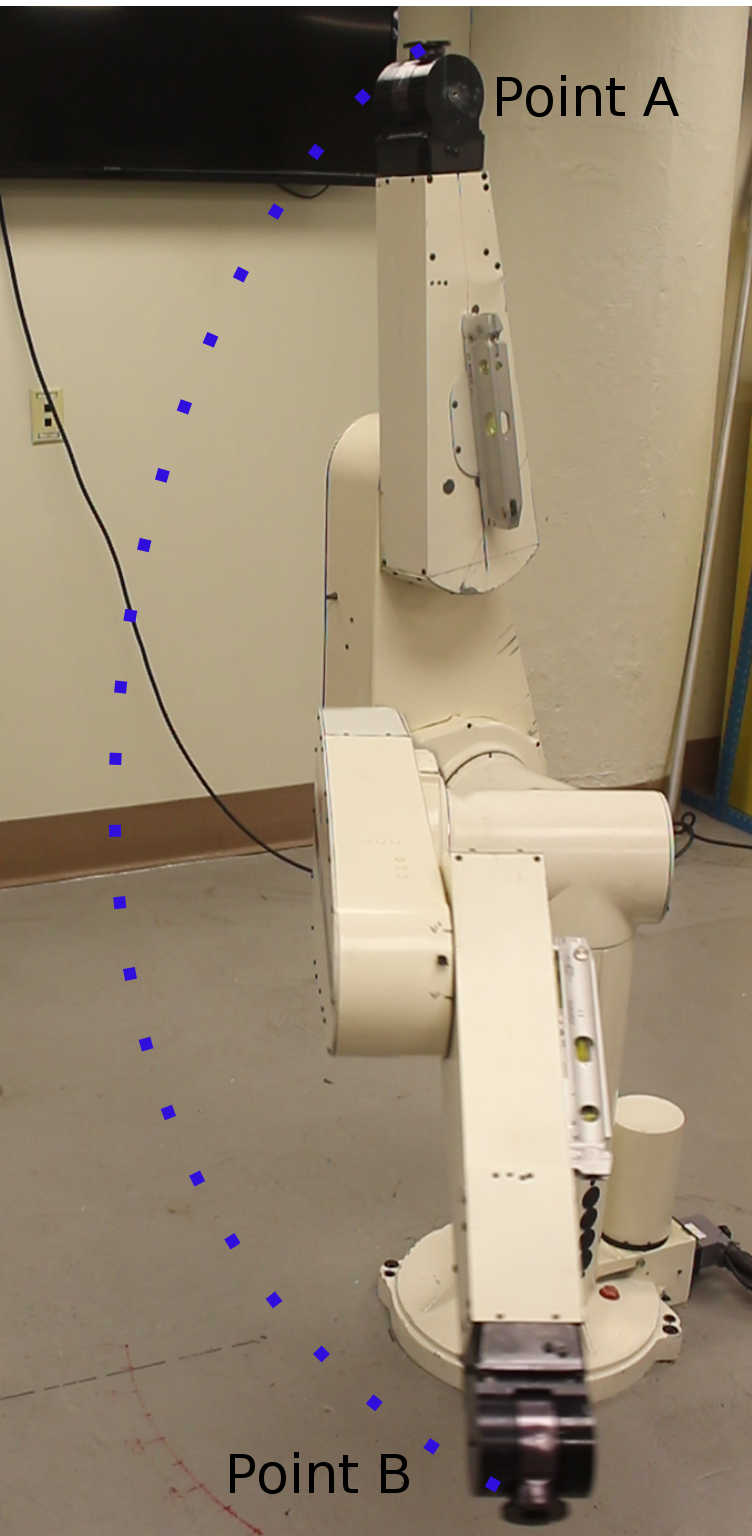}
\caption{}
\label{puma2}
\end{subfigure}
\caption{The PUMA 560 robot used as a case study for finding optimal trajectories maximizing energy regeneration, a) coordinate frames for modeling the PUMA robot assigned using the Denavit-Hartenberg convention, b) the starting position, referred to as point A, and the final position, referred to as point B.}
\label{puma} 
\end{figure}

As a case study, we consider finding optimal trajectories for a PUMA 560 robot. However, the methods used here are applicable to any robotic manipulator that can be modeled as in Eq.\eqref{augmodel}. The PUMA robot shown in Fig.~\ref{puma1}, consists of three main joints and spherical wrist, which together provide six degrees of freedom for the robot. Here, we only consider the dynamics of the three main joints of the robot which have the most potential for energy regeneration. 
The three main joints, $q_1$, $q_2$, and $q_3$, are assumed to be semi-active and connected in the star configuration via a central ultracapacitor. The robot is constrained to start from the initial position $q_{start}=[0,-\pi/2,0]$ and initial velocity $\dot{q}_{start}=[0,0,0]$ -- referred to as point A -- and finish at $q_{end}=[\pi/3,0,\pi/4]$ with $\dot{q}_{end}=[0,0,0]$ -- referred to as point B. Point A and B are shown in Fig.~\ref{puma2}. Note that point A is at a higher potential energy level compared to point B.
Using the linear parameterization property for robotic manipulators~\cite{SHV}, and assuming no external forces and moments are applied to the robot ($\mathcal{T}=0$), the dynamic equations for the PUMA robot (Eq.~\eqref{consta}) can be written as
\begin{equation}
u=Y(q,\dot{q},\ddot{q})\theta
\label{augmodelreg}
\end{equation}
where $Y_{n\times p}$ is the regressor matrix, and $\theta_{p \times 1}$ is the parameter vector which is a function of all the physical parameters of the system (e.g. link lengths, link masses, gear ratios etc.). Using the Denavit-Hartenberg (DH) convention \cite{SHV}, dynamic equations for the PUMA robot and the semi-active drive mechanisms are derived. These equations are presented in regressor and parameter vector form in Appendix~\ref{appen}. Figure~\ref{puma1} shows coordinate frames assigned for the PUMA robot using the DH convention.   

The optimal control problem defined in Eq.~\eqref{costfunc} and Eq.~\eqref{const} is in general nonlinear and non-convex. It can have none, one, many, or an infinite number of solutions. In most cases, no immediate analytical solution exists and one normally resorts to numerical methods for solving the problem. The optimality conditions for this problem generally lead to a set of differential equations with split boundary conditions. Methods such as steepest decent and variation of extremal are are used for solving theses types of two point boundary value problems \cite{kirk2012optimal}.

After deriving dynamic equations for the PUMA robot and the regenerative semi-active joints, we use the method of direct collocation \cite{von2013numerical,van2011implicit,rohani2017optimal} to transcribe the optimal control problem  into a large-scale nonlinear program (NLP) problem. In this method, the states ($q$, $\dot{q}$) and controls ($\tau^d$) are discretized into into $N$ temporal nodes. The cost function (Eq.~\eqref{costfunc}) and constraints (Eq.~\eqref{const}) are discretized by using an appropriate finite difference approximation for the state derivatives. We use the backward Euler approximation in this work. The cost function becomes a function of the states and controls at each grid point, and the dynamic constraints are converted into a set of algebraic constraints that are also a function of the discretized states and controls. The optimal control problem is converted to a constrained optimization problem of finding the states and controls at each grid point that minimize the discretized cost function and satisfy a set of algebraic constraints.

\section{Numerical Optimization Results}\label{sec:optres}
The code used to solve the problem considered in this paper is available~\cite{code}. The direct collocation problem is solved using the IPOPT (interior point optimizer) numerical solver~\cite{wachter2006implementation}. The IPOPT solver generally finds local optima for nonlinear problems. To find the global optimum, the optimization is run several times starting from different random initial conditions. By doing this one can have a practical assurance that the problem has converged to the global optimum.  For our problem, all the initial conditions tested converged to the same optimal solution. In addition, using successive mesh refinement, the value of $N=100$ was found for which the results of the optimization showed little variation with respect to the value of $N$.

\begin{figure}[!t]
\centering
    \begin{subfigure}[t]{0.5\textwidth}
        \centering
        \includegraphics[width=2.5in]{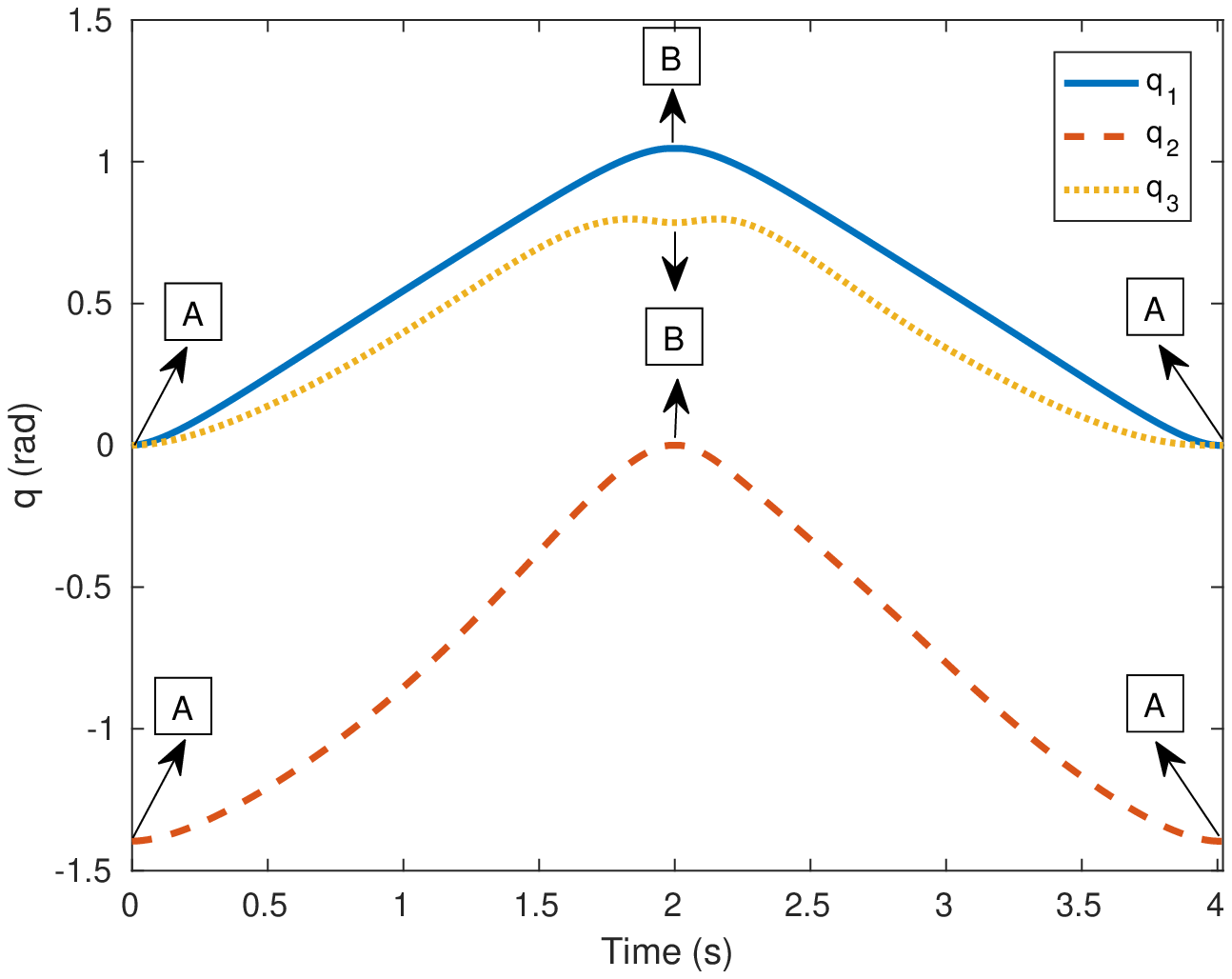}
        \caption{}
    \end{subfigure}
    \begin{subfigure}[t]{0.5\textwidth}
        \centering
        \includegraphics[width=2.5in]{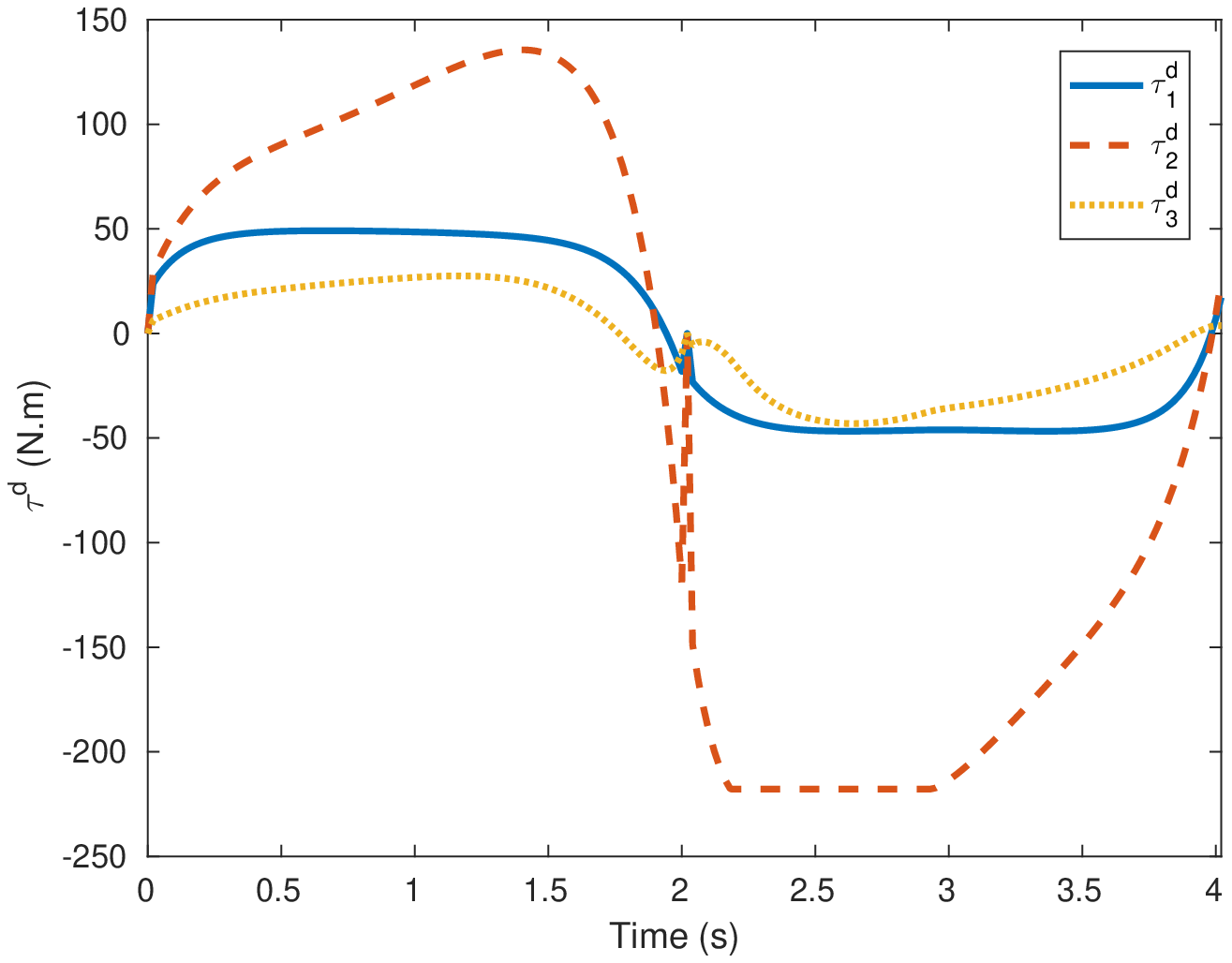}
        \caption{}
        \label{optimaltraj-2}
    \end{subfigure}    
\caption{Optimal point-to-point trajectories and controls found for the PUMA 560 robot, a) optimal trajectories from point $A$ (starting point) to point $B$ (final point), and from point $B$ to point $A$, b) the optimal controls ($\tau^d$) that will results in the optimal trajectories (controls are bound between $[-135.51,135.51]$ Nm, $[-217.85,217.85]$ Nm and  $[-117.67,117.67]$ Nm for Joints~1, 2, and 3 respectively).}
\label{optimaltraj} 
\end{figure}

The optimization was run once from a starting point $A$ to the final point $B$, and once from $B$ back to $A$. Figure \ref{optimaltraj} shows the optimal trajectories and controls found. Note that with an initial capacitor voltage of $27$ Volts, the controls bounds were calculated as  $[-135.51,135.51]$ Nm, $[-217.85,217.85]$ Nm and  $[-117.67,117.67]$ Nm for Joints~1, 2, and 3 respectively.

\begin{figure}[!t]
\centering
\includegraphics[width=2.5in]{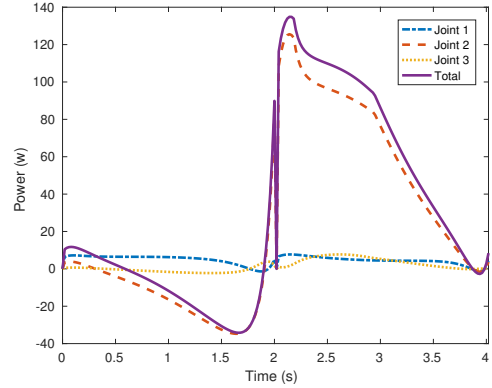}
\caption{Theoretical power flows resulted when following the optimal trajectories for the PUMA 560 robot. Figure shows power flow for Joints~1 to 3 and also the total power flow (sum of power flows, total power from the capacitor to the robot). Power is positive when going from the capacitor to the robot joints.}
\label{optimalTHpower} 
\end{figure}

Figure~\ref{optimalTHpower} shows the theoretical power flows that would result if the robot followed the optimal trajectories. Power is positive  when it flows from the capacitor to the motor of each joint. The total power flow (the sum of power flows) represents the power flow from the capacitor to the robot.

Table~\ref{table1} compares the energy consumption of each joint when going from $A$ to $B$ and viceversa. Positive energy indicates that energy is consumed, negative energy indicates that energy is regenerated. From this table and Fig.~\ref{optimalTHpower} it is seen that when going from $A$ to $B$, Joints~2~and~3 are regenerating, energy while Joint~1 is consuming energy. Form $B$ to $A$, however, all joints are consuming energy. These results are somewhat expected since at point $B$, Joints~2 and~3 are at a lower potential energy level compared point $A$. Therefore, the potential energy difference between points $A$ and $B$ is being regenerated and partially stored in the common capacitor. It is also observed that Joint~2 is the main contributor to energy regeneration when going from $A$ to $B$ due to motion in a vertical plane and a large weight. When going from $B$ to $A$, the capacitor needs to provide the potential energy difference between the two points to move the robot back to point $A$.
\begin{table}[!b]
\centering
\caption{Theoretical energy consumption for each joint of the PUMA 560 robot when the joints follow optimized trajectories. Positive energy indicates  energy consumption, negative energy indicates energy regeneration.}
\ra{1.3}
\begin{tabular}{@{} l c c  c @{}}\toprule
  && $E_{A\to B } (J)$ & $E_{B\to A } (J)$\\
  \midrule
  Joint 1&& 9.94  & 9.29 \\
  Joint 2&& -27.31& 96.17 \\
  Joint 3&& -1.96 & 10.15 \\
  Total  && -19.33& 115.61\\
  \bottomrule
\end{tabular} 
\label{table1}
\end{table}

In Section \ref{sec:expeval}, the optimal trajectories are implemented on the PUMA 560 robot to experimentally evaluate energy regeneration.

\section{Experimental evaluation}\label{sec:expeval}
\begin{figure*}[!t]
\centering
\includegraphics[width=5in]{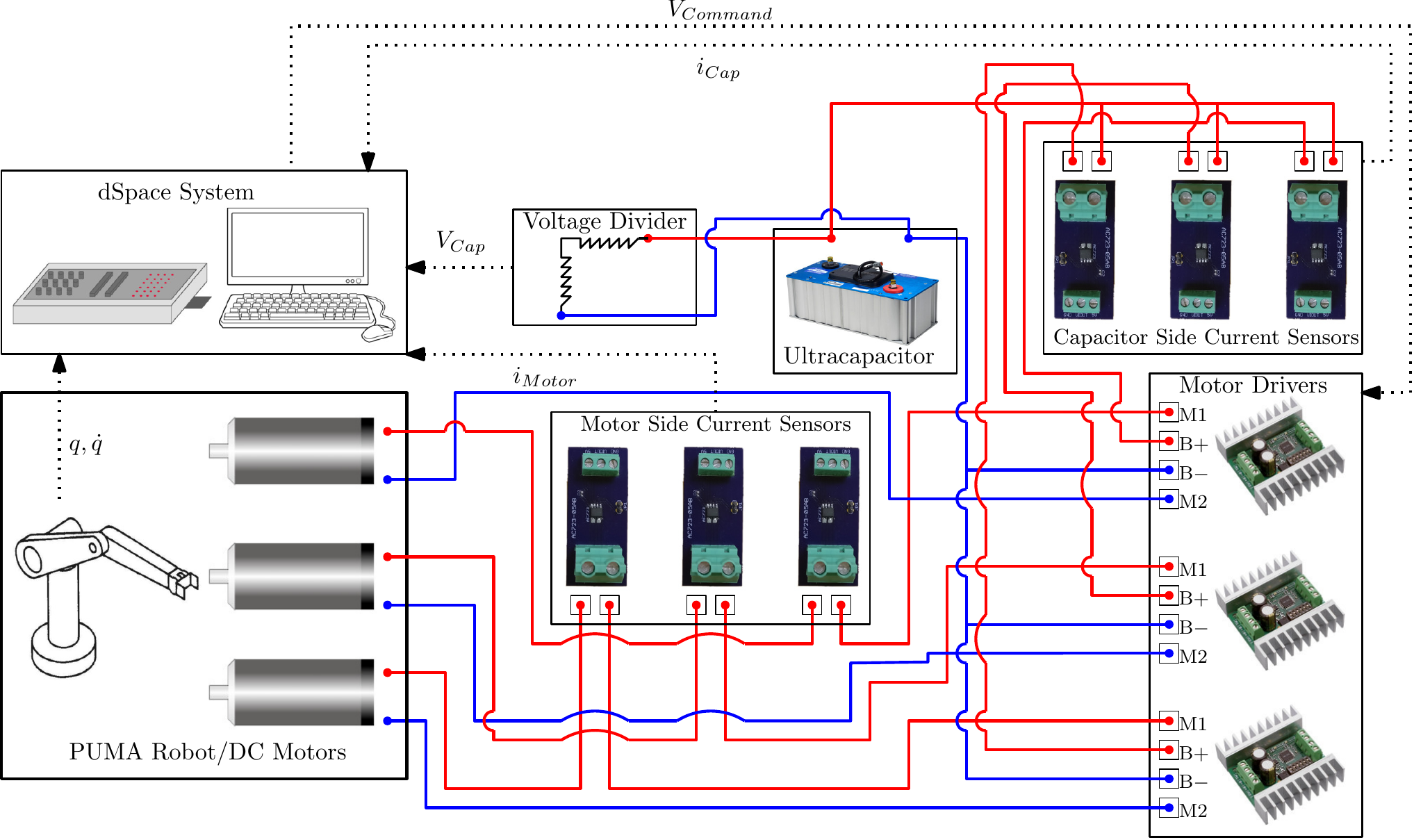}
\caption{Schematic of the experimental setup. Current sensors are used to measure currents on both sides of each motor driver. Voltage on the capacitor side is measured  directly via the dSpace system. The dSpace system is also used for controlling the robot. The motor drivers are all connected to a common ultracapacitor (star configuration). Dotted lines indicate signals, solid lines indicate wiring.}
\label{expschem} 
\end{figure*}

Figure \ref{expschem} shows the schematic of the experimental setup. The PUMA 560 robot has three main joints and three joints at the wrist. Here we are only concerned with the operation of the main three joints. Each joint is actuated by a DC motor that is driven by the four quadrant 25 amp SyRen motor driver (Dimension Engineering, Hudson, Ohio). The dSPACE 1103 controller board (dSPACE GmbH, Paderborn, Germany) is used for controlling the robot and for data acquisition. The input and output voltages and currents for each motor driver are needed to calculate the instantaneous power. The currents are measured via the ACS723 current sensors (Allegro Microsystems, Worcester, Massachussetts). The capacitor voltage is directly measured by using a voltage divider and the dSPACE system. The voltage on the motor side is not directly measured, however it is verified separately that this voltage accurately follows the voltage requested by the command signal~($V_{Command}$). The input of all three motor drivers are connected to a common 48 V ultracapacitor bank (Maxwell Technologies, San Diego, California) with a capacitance of $165$ F. The capacitor is initially charged to $27$ volts to avoid reaching the $30$ volts absolute maximum input voltage for the motor drivers. A robust passivity-based control method is used to track the optimal trajectories found in Section~\ref{sec:optres}. The controller is implemented in real time with the dSPACE system and uses angular position and velocity feedback provided by encoders on the robot joints, in addition to capacitor voltage feedback. Figure~\ref{expschem} shows the experimental setup.

\subsection{Overview of robust passivity-based control}\label{RPBC}
The optimization problem yields an open loop solution which is not implementable in the real robot. The robust passivity-based control~\cite{SHV} was selected to ensure the robot tracks the desired optimal trajectories with guaranteed stability against parametric uncertainties in the robot model. Based on the dynamic equation for the augmented model (Eq.~\eqref{augmodel}) and assuming no known external forces or moments are exerted on the robot ($\mathcal{T}=0$), the control input is chosen as
\begin{equation}
\tau^d=\hat{D} a+\hat{C}\nu +\hat{\mathcal{R}}+\hat{g}-Kr=Y_a(q,\dot{q},a,v)\hat{\theta}-Kr
\end{equation}
where $Y_a$ is the control regressor and $\hat{\theta}$ is the parameter estimate adjusted by the control law. Variables $a$, $v$, and $r$ are defined as
\begin{subequations}
\begin{align}
&v=\dot{q}^d-\Lambda \tilde{q}\\
&a=\dot{\nu}\\
&r=\dot{q}^d-\nu
\end{align}
\end{subequations}
where $q^d$ and $\dot{q}^d$ denote the desired joint trajectories and $\tilde{q}=q-q^d$ denotes the tracking error. Also, $K$ and $\Lambda$ are diagonal matrices with positive nonzero entries. The parameter estimate $\hat{\theta}$ is adjusted according to
\begin{equation}
 \hat{\theta}=\theta_0+\delta \theta
\end{equation}
where $\theta_0$ is a set of constant nominal parameters. if the parametric uncertainty of the system is bounded, $\| \theta -\hat{\theta}\| \leq \rho$, then choosing $\delta \theta$ as
\begin{equation}
\delta \theta=
\begin{cases}
-\rho \frac{Y_a^T r}{\|Y_a^T r\|}       & \quad \text{if } \|Y_a^T r\| > \epsilon\\
    -\frac{\rho}{\epsilon} Y_a^T r      & \quad \text{if } \|Y_a^T r\| \leq \epsilon
\end{cases}
\end{equation}
where $\epsilon$ is a small positive parameter, one can show ultimate boundedness of the tracking error. 

\subsection{Experimental results}
\begin{figure}[!t]
\centering
    \begin{subfigure}[t]{0.5\textwidth}
        \centering
        \includegraphics[width=2.5in]{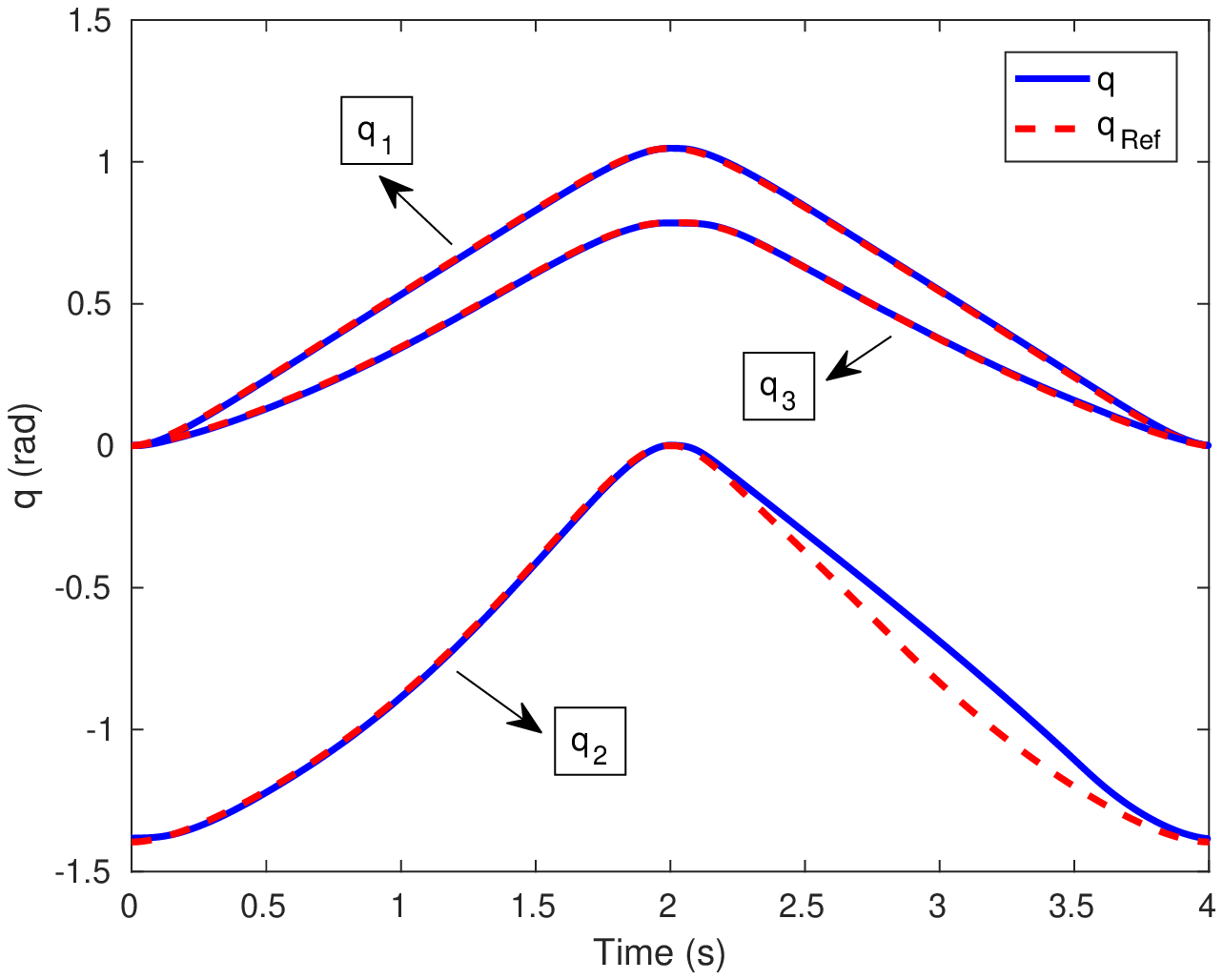}
        \caption{}
        \label{actual&desiredtraj-1}
    \end{subfigure}
    \begin{subfigure}[t]{0.5\textwidth}
        \centering
        \includegraphics[width=2.5in]{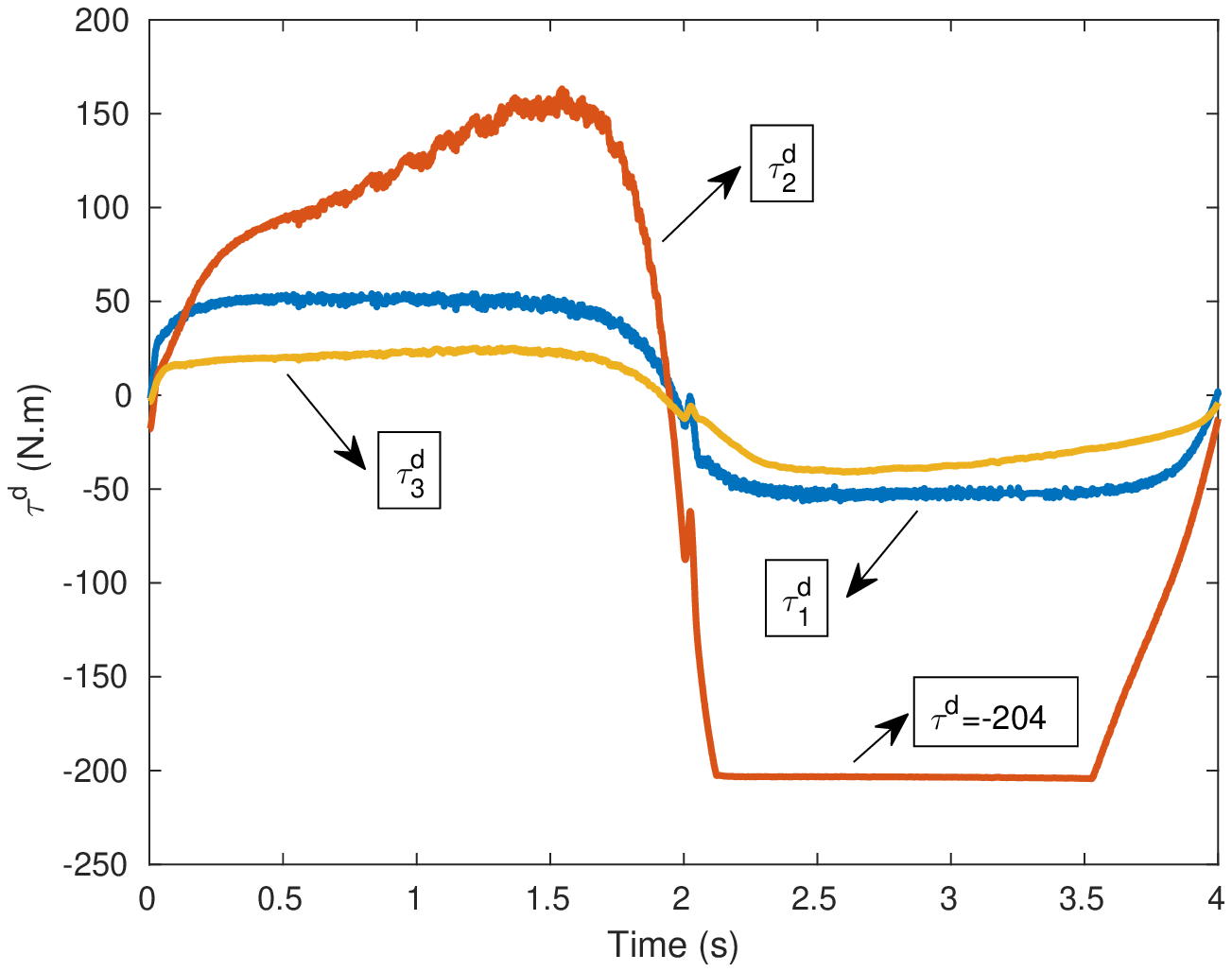}
        \caption{}
        \label{actual&desiredtraj-2}
    \end{subfigure}    
\caption{a) Actual and optimal reference trajectories for the PUMA 560 robot, b) The control commands ($\tau^d$) that results in the trajectories. The robust control method used provides good tracking of the optimal reference trajectories.}
\label{actual&desiredtraj} 
\end{figure}

\begin{figure}[!t]
\centering
\includegraphics[width=2.5in]{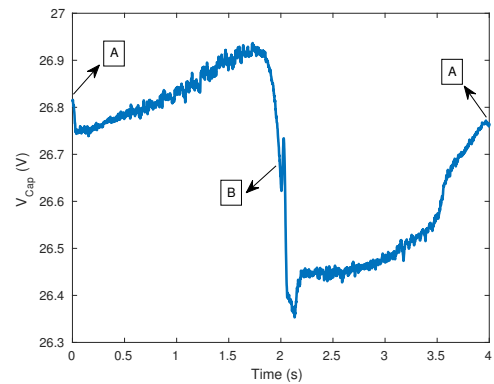}
\caption{Capacitor voltage during the movement of the robot from Point $A$ to Point $B$ and vice versa. The capacitor voltage starts from $26.82$ volts at the beginning of the movement and ends up with $26.76$ volts at the end of the movement.}
\label{Vcapexp} 
\end{figure}

Figure~\ref{actual&desiredtraj-1} shows the actual and optimal reference trajectories followed by the robot joints. We see that the robust passivity based controller provides very good tracking of the optimal reference trajectories. The controller does lose tracking to a small degree for the second joint when going form point $B$ to $A$ (maximum error is $0.12$ radians). The cause of this loss of tracking can be found by observing the control inputs $\tau^d$, Fig.~\ref{actual&desiredtraj-2}. The control input for the second joint ($\tau_2^d$) saturates around $-204$ N.m, which is higher compared to the saturation value for the optimum control ($-217.85$ N.m, Fig.~\ref{optimaltraj-2}). The $-217.85$~N.m bound for the optimization was set assuming a constant capacitor voltage of $27$ volts. However, the capacitor voltage does not stay constant during the movement of the robot as seen in Fig.~\ref{Vcapexp}. The capacitor voltage is about $26.82$ volts at the beginning of the movement and varies between $26.94$ and $26.35$ volts. During the $B$ to $A$ portion of the movement, the capacitor voltage is less than $27$ volts, therefore, there is not enough voltage in the capacitor to follow the optimum trajectory for Joint~2. More accurate results can be achieved by including the ultracapacitor model in the optimization, however doing so would significantly increase the complexity of the problem. 

\begin{figure}[!t]
\centering
    \begin{subfigure}[t]{0.5\textwidth}
        \centering
        \includegraphics[width=2.5in]{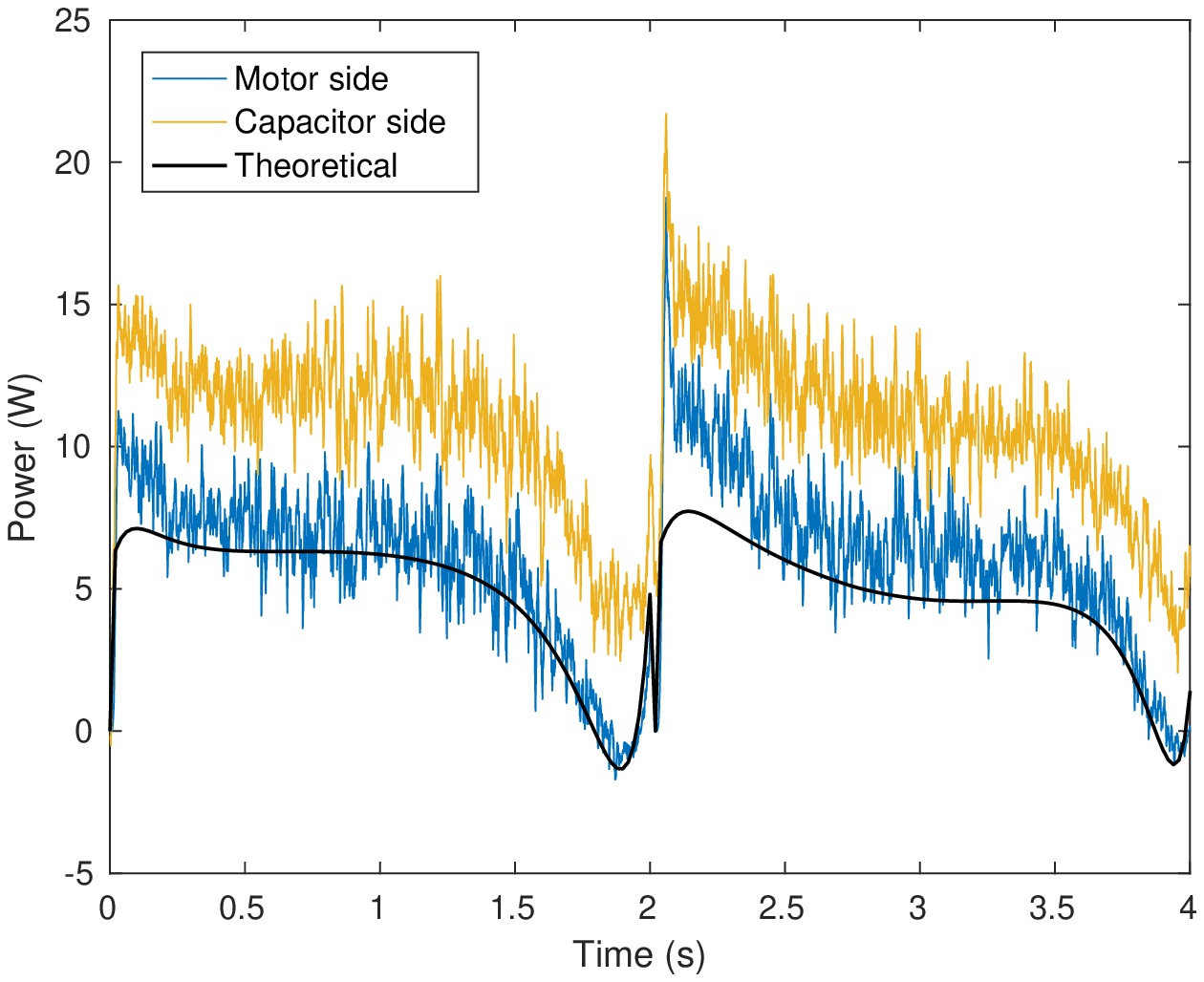}
        \caption{}
    \end{subfigure}
    \begin{subfigure}[t]{0.5\textwidth}
        \centering
        \includegraphics[width=2.5in]{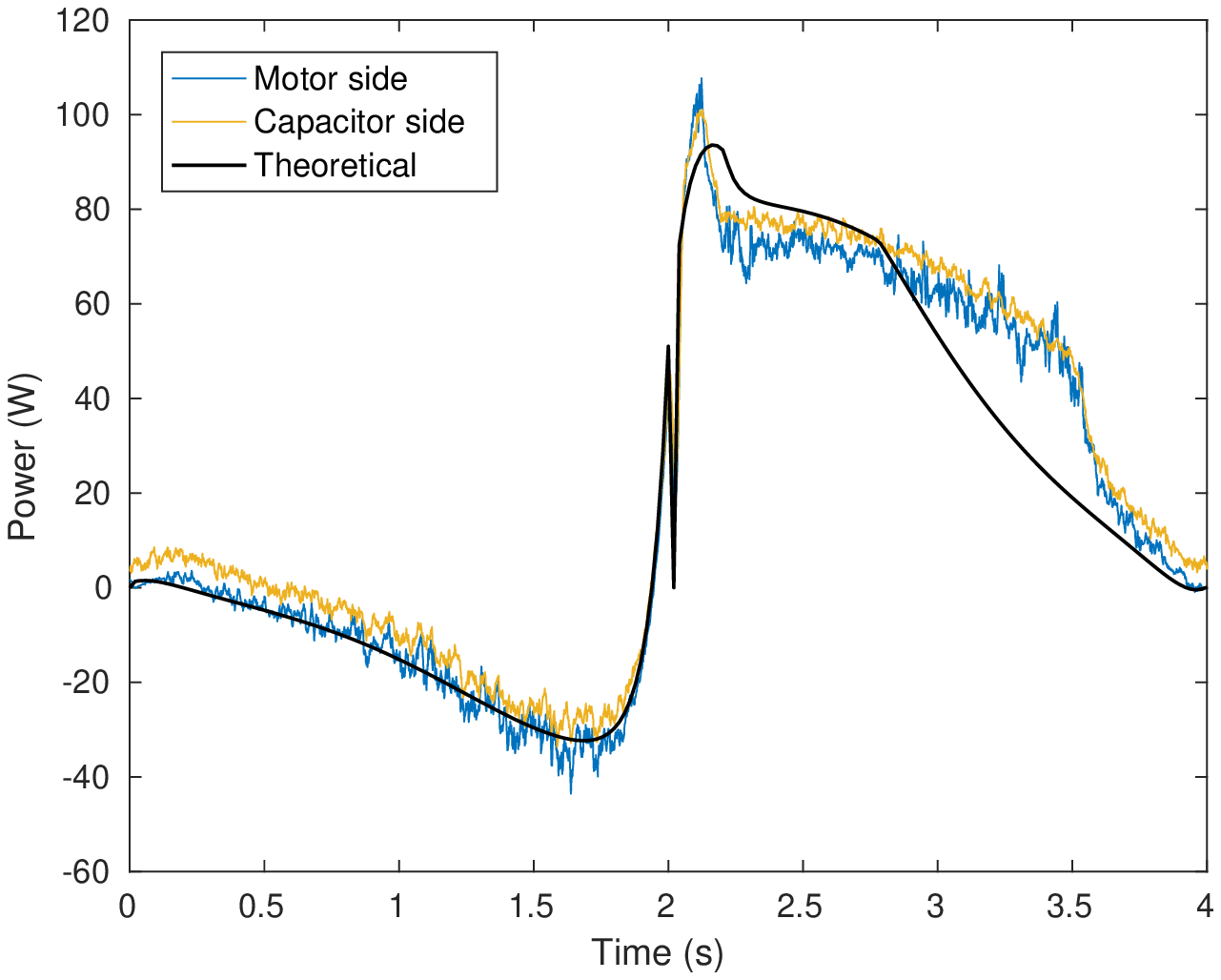}
        \caption{}
    \end{subfigure}
    \begin{subfigure}[t]{0.5\textwidth}
        \centering
        \includegraphics[width=2.5in]{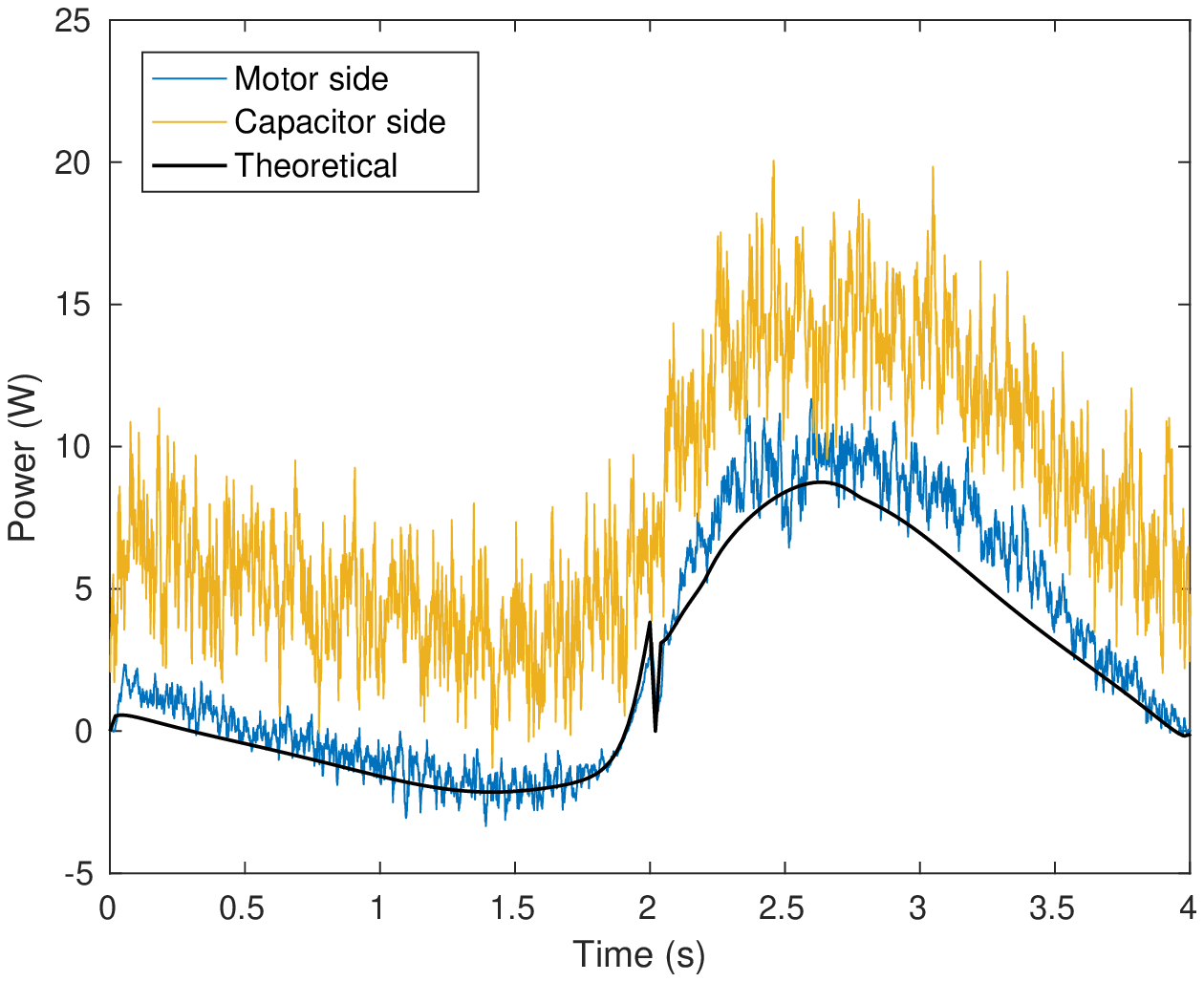}
        \caption{}
    \end{subfigure}
\caption{Power flows for the motor side and capacitor side of the motor driver for a) Joint~1, b) Joint~2 and c) Joint~3. Positive power indicates energy consumption and negative power indicates energy regeneration. The theoretical power flow is also shown for comparison.}
\label{powerflow} 
\end{figure}

Figure~\ref{powerflow} shows power flows for the motor side and capacitor side of the motor drivers. Power is positive when it flows from the ultracapacitor to the motor driver and from the motor driver to the robot joints. Figure \ref{powerflow} also compares the theoretical power flows found from the optimization (also shown in Fig.~\ref{optimalTHpower}) with the experimental power flows. For Joint~1, power on the motor side is always positive, meaning that there is no energy regeneration. This agrees with what was found for the theoretical power. In general, the theoretical power for Joint~1 agrees quit well with the power on the motor side. The relatively small disagreement in the $B$ to $A$ portion could be due to underestimation of friction in the model (more power is required to actuate the joint than calculated). Results also show that the power on the capacitor side is higher than the power on the motor side. This reflects the inefficiency of the motor driver (some power is dissipated in the motor driver) and also the power required to operate them. These inefficiencies are not taken into account by the optimization. Joint~2 shows a significant amount of negative power in the first portion of the trajectory on the motor side of the motor drive (energy is being regenerated). The negative peak power on the motor side is $-44$ watts. In this portion, part of the power regenerated on the motor side reaches the capacitor and the rest is dissipated in the motor driver. This is inferred  from the fact that power on the capacitor side is less negative compared to the motor side. The theoretical power agrees very well with the power on the motor side for the first portion of the trajectory. The discrepancy for the second portion of the trajectory can be related to the loss of tracking in Joint~2. In the second portion of the trajectory, power is positive and energy is being consumed. Power on both sides of the motor driver coincide indicating that the motor driver is in its efficient operating range. Joint~3 shows portions of negative power on the motor side however power on the capacitor side is only positive. This indicates that energy is being regenerated but it is all dissipated in the motor driver and does not reach the capacitor. The theoretical power again matches very well with the motor side power. The agreement between theoretical and actual power flows show that the model was obtained with good accuracy. Also, note that that in portions of the robot's movement, Joint~2 is regenerating energy while Joints~1 and Joint~2 are consuming energy. Thus energy is channeled from Joint~2 to the other robot joints 
through the capacitor.

Note that the dynamic behavior of ultracapacitors varies from conventional capacitors. Ultracapacitors are nonlinear due to the capacitance being a function of voltage and frequency \cite{grbovic2013ultra,musolino2013new}. A consequence of this is that an increase in ultracapacitor voltage does not necessarily indicate an increase in stored energy (i.e., $E=CV^2/2$ does not hold). This can be clearly observed from Fig.~\ref{Vcapexp} where in the $B$ to $A$ portion, ultracapacitor voltage is increasing in spite of energy being consumed by the robot (Fig.~\ref{powerflow}). By using the capacitor voltage feedback (Eq.~\eqref{taud}), SVC allows the control of the robot without being concerned with the nonlinearities of the ultracapacitor and their effect on the overall behavior of the robot.

Integrating the power flows over time yields the energy consumption for the motor side and capacitor side of the motor driver. These results are summarized in Table~\ref{table2} along with a comparison to the theoretical energy consumption. It is seen that the theoretical values agree relatively well with the experimental values for the motor side. The agreement worsens on the capacitor side due the unmodeled  efficiency of the motor driver. For Joint~2, the theoretical model, compared to the experimental results, regenerates less energy when going from $A$ to $B$ and consumes less energy when going from $B$ to $A$. Although small, this effect reflects a discrepancy between the model and the robot and small errors in measurements. Only part of the energy that reaches the motor driver is stored in the capacitor. Dividing the capacitor side energy by the motor side, the regeneration efficiency of the motor driver for Joint~2 is about $65\%$.
\begin{table*}[!t]
\centering
\caption{Energy consumption for the PUMA 560 robot when following optimal trajectories. Energy consumption is reported for the motor side and capacitor side of the motor driver, when going from point $A$ to point $B$ and vice versa. Theoretical results are also reported for comparison. Negative energy indicates energy being regenerated.}
\ra{1.3}
\begin{tabular}{@{} l c c c c c c c c @{}}\toprule
  && \multicolumn{3}{c}{$E_{A \to B} (J)$} & \phantom{abc} & \multicolumn{3}{c}{$E_{B \to A} (J)$}\\
  \cmidrule{3-5}  \cmidrule{7-9}
  && Motor & Capacitor& Theoretical & & Motor & Capacitor& Theoretical \\
   \midrule
  Joint 1 &&  11.43& 21.27&  9.94  && 12.52 & 21.62 & 9.29   \\
  Joint 2 && -27.76&-18.16& -27.31 && 106.16& 113.93& 96.17 \\
  Joint 3 && -1.18 & 9.23 & -1.96  && 12.61 & 22.87 & 10.15   \\
  Total   && -17.51&12.34 & -19.33 && 131.29& 158.42& 115.61\\
  \bottomrule
\end{tabular} 
\label{table2}
\end{table*}

\begin{figure}[!t]
\centering
    \begin{subfigure}[t]{0.5\textwidth}
        \centering
        \includegraphics[width=2.5in]{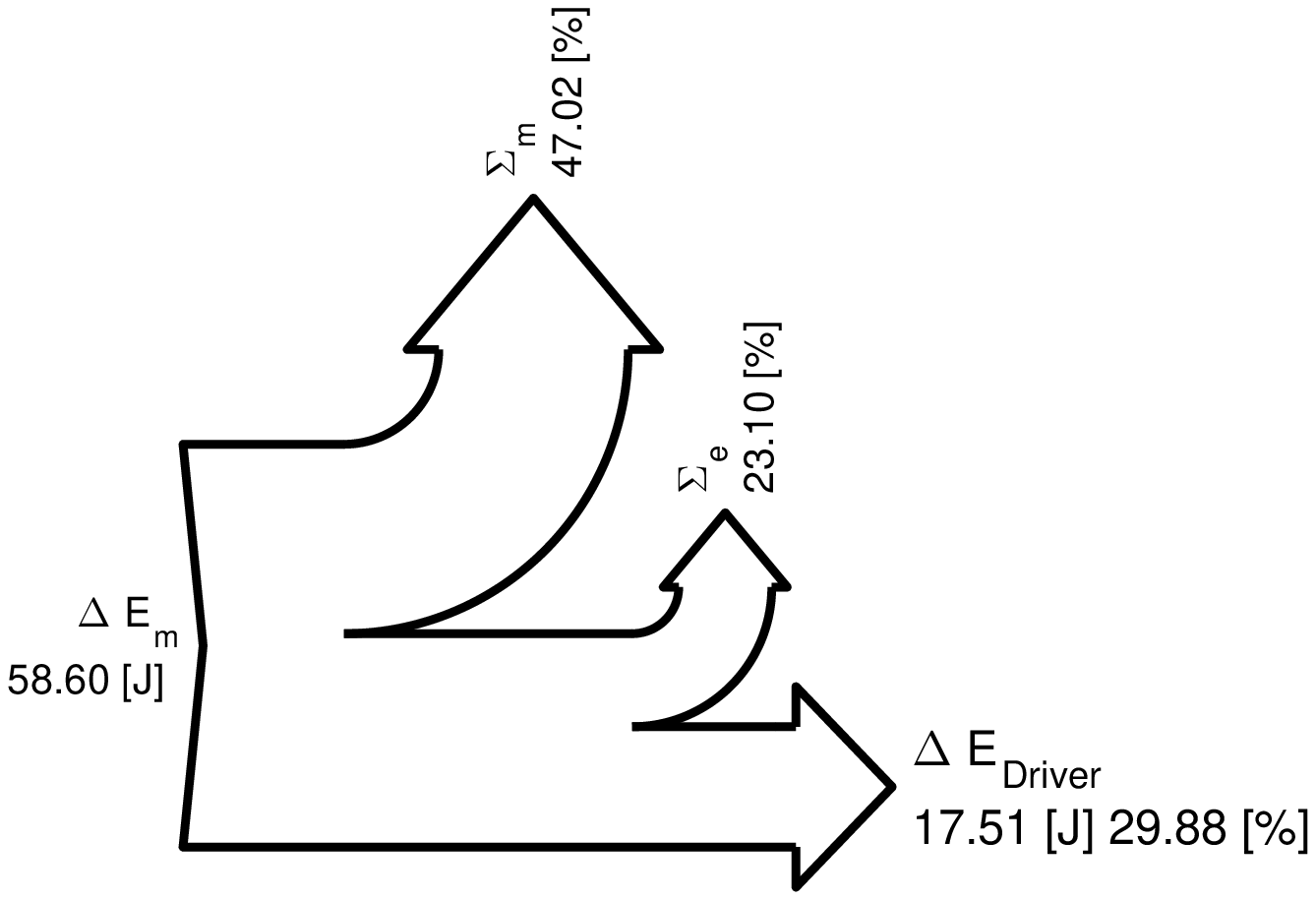}
        \caption{}
    \end{subfigure}
    \begin{subfigure}[t]{0.5\textwidth}
        \centering
        \includegraphics[width=2.5in]{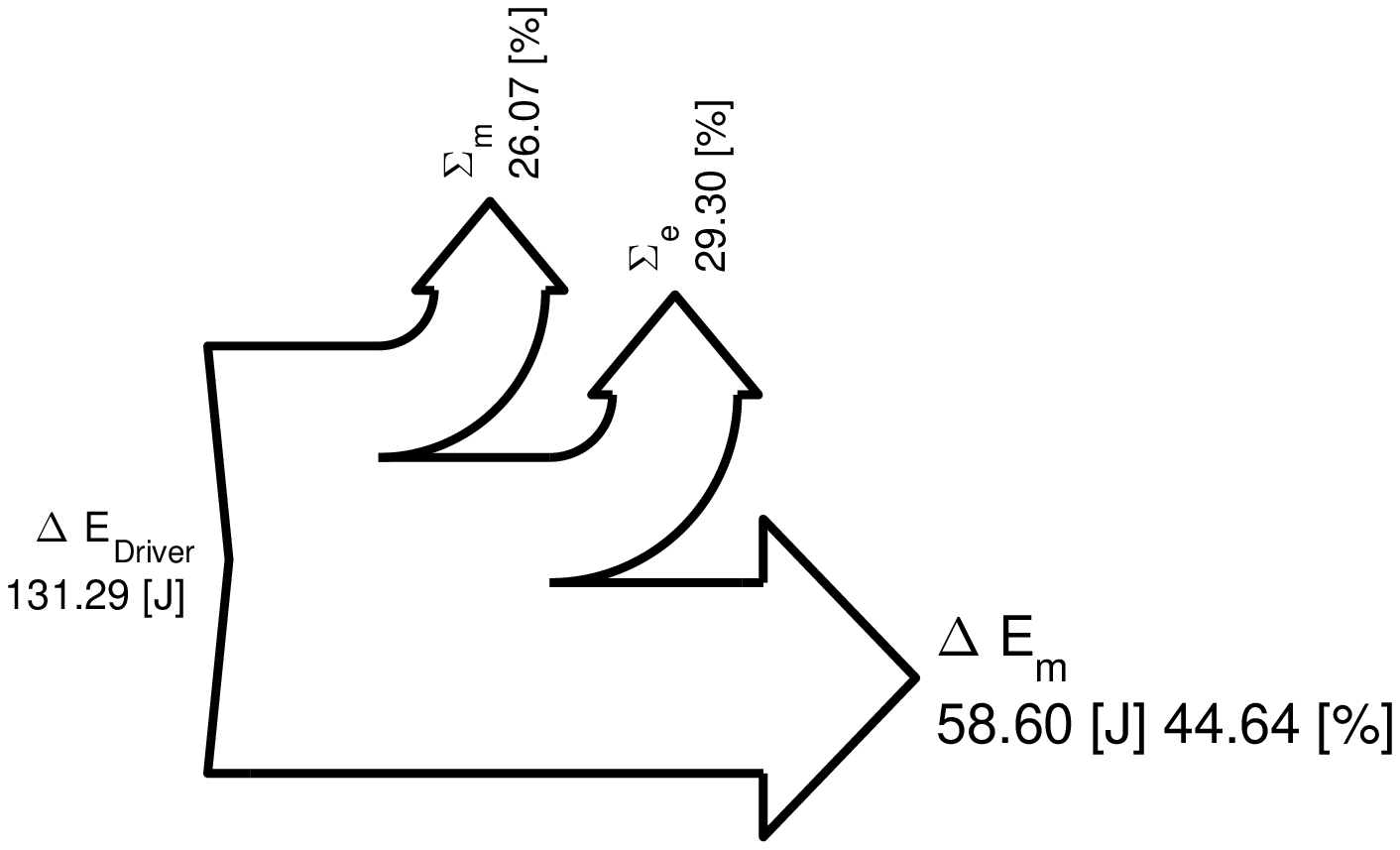}
        \caption{}
    \end{subfigure}
\caption{Sankey diagram showing the overall energy balance for the PUMA 560 robot when following optimal trajectories form a) $A$ to $B$ and b) $B$ to $A$. The overall mechanical energy of the robot is represented by $\Delta E_m$, $\Sigma_e$ and $\Sigma_m$ are the electrical and mechanical losses respectively, $\Delta E_{Driver}$ is the energy going to (i.e. regenerated) or coming from (i.e. consumed) the motor driver.}
\label{sankey} 
\end{figure}

Figure~\ref{sankey} shows Sankey diagrams for the overall energy balance based on Eq.~\eqref{extenergy} and using model parameters. Since there are no external forces or moments applied to the robot (i.e. $W_{ext}=0$) and all the robot joints are semi-active (i.e. $W_{act}=0$), energy can be stored in the capacitor only due to changes in mechanical energy ($\Delta E_m^T$). The difference in mechanical energy between points $A$ to $B$ is $58.6$ J. In the  first portion of the movement,  about $47\%$ of the mechanical energy is dissipated as mechanical losses, another $ 23\%$ is dissipated as electrical losses, and only about $30\%$ reaches the motor drive. Due to inefficiencies in the motor drive only part of that energy is actually stored in the capacitor; however by utilizing a high efficiency drive these additional losses can be minimized. In the second portion of the movement, $131.29$ J of energy is provided by the driver to move the robot from point $B$ to point $A$. Mechanical losses account for about $26\%$ of the provided energy, electrical losses account for about $30\%$ of the provided energy, and only $44\%$ is stored as mechanical energy. These figures indicate that the mechanical losses, which are due to the design of the robot, are a large portion of the total losses and a better robot design can lead to more energy regeneration. In the total cycle, $131.29$ J of energy was provided by the motor drive in which $17.51$ J was regenerated. Therefore energy regeneration resulted in about $13\%$ reduction in the total energy consumed.

\begin{figure}[!t]
\centering
   \begin{subfigure}[t]{0.5\textwidth}
       \centering
       \includegraphics[width=2.5in]{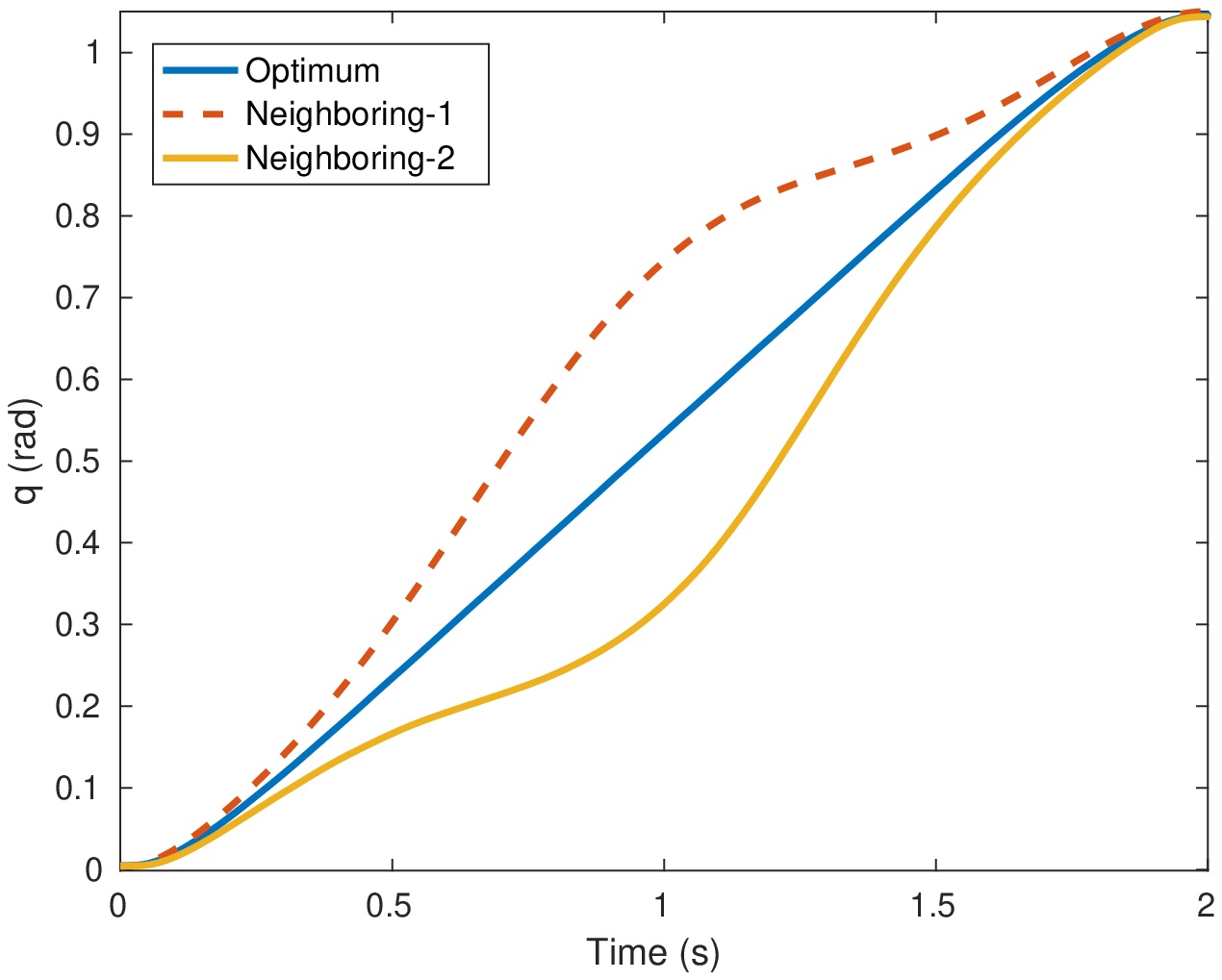}
       \caption{}
   \end{subfigure}
   \begin{subfigure}[t]{0.5\textwidth}
       \centering
       \includegraphics[width=2.5in]{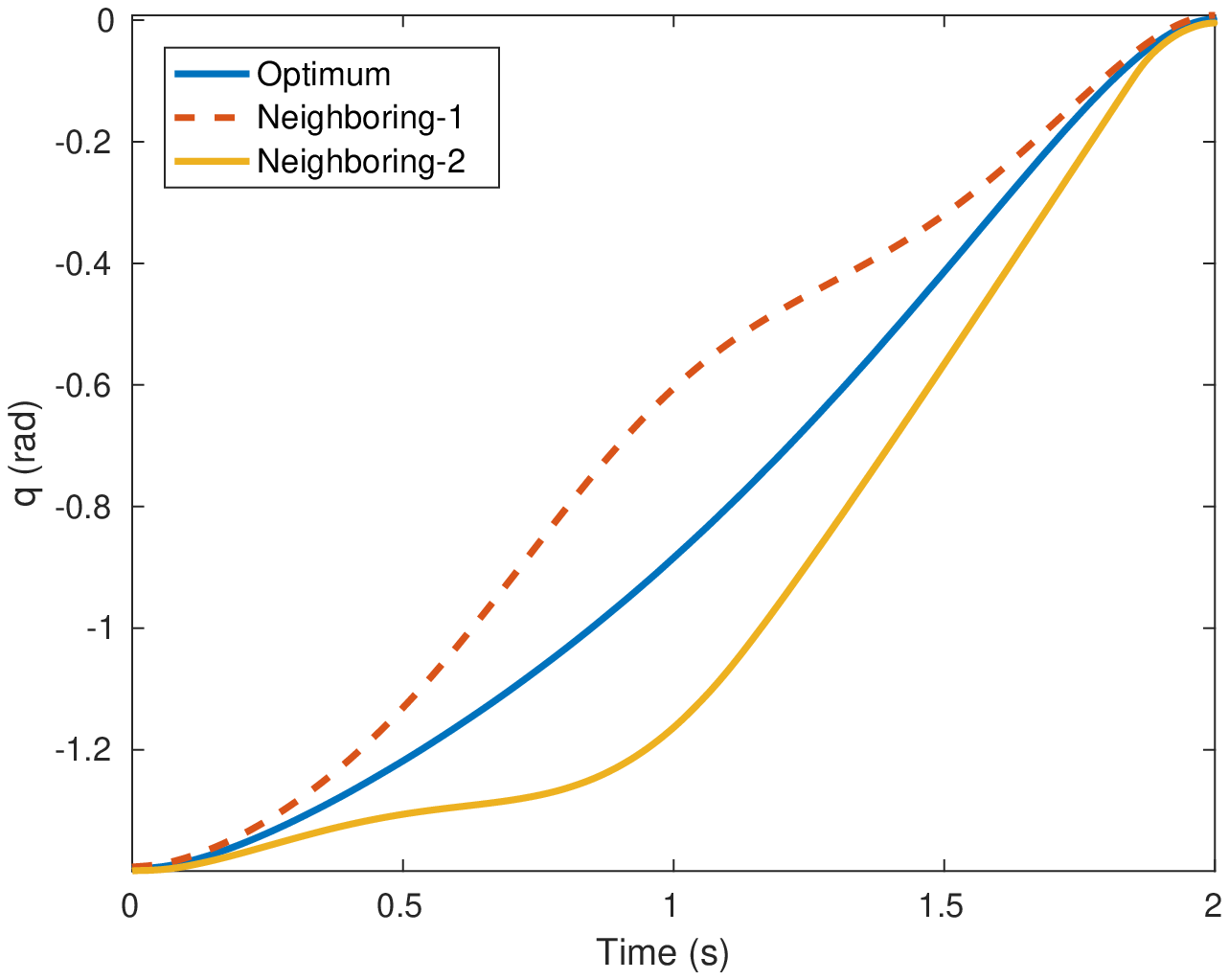}
       \caption{}
   \end{subfigure}
   \begin{subfigure}[t]{0.5\textwidth}
       \centering
       \includegraphics[width=2.5in]{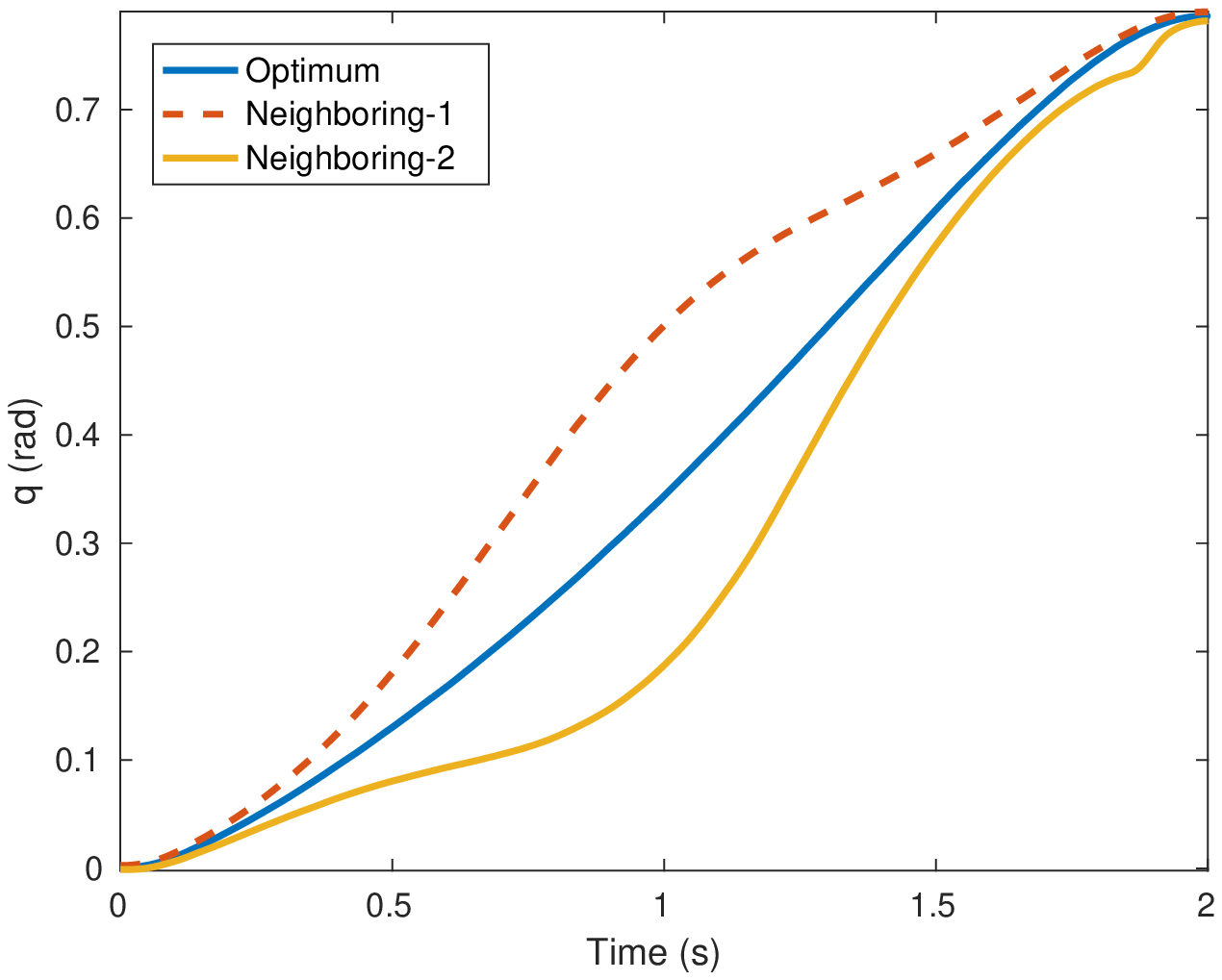}
       \caption{}
   \end{subfigure}
\caption{Optimum and neighboring trajectories followed by the robot for a) Joint~1, b) Joint~2, and c) Joint~3, when going from point $A$ to point $B$. Neighboring trajectories are tested to show the effectiveness of the optimization}
\label{neighbortraj} 
\end{figure}

To verify that the optimum trajectories are in fact maxima, two neighboring trajectories are generated and evaluated. In the interest of conciseness, we only consider the first portion of the movement (from $A$ to $B$). Neighboring trajectories were generated by adding a Gaussian function term to the optimum trajectory
\begin{equation}
q_{neighboring}=q_{optimum}\pm\varepsilon \mathrm{e}^{-\frac{1}{2}\left(\frac{(t-\mu)^2}{\sigma^2}\right)}
\end{equation}
With $\mu=1$, $\sigma=\mu /3$, and $\varepsilon=0.2\max(|q|)$. Parameters for the Gaussian function are chosen so that the neighboring trajectories satisfy the boundary conditions for the optimal trajectory with negligible error. Figure~\ref{neighbortraj} shows the neighboring trajectories followed by the robot. Table~\ref{table3} compares energy consumption for the optimum trajectory with that of the neighboring trajectories. We see that even though the optimum trajectory consumes slightly more energy for Joint~1, the total energy regenerated is higher for the optimum trajectory.

\begin{table}[!t]
\centering
\caption{Comparison of energy consumption for the PUMA 560 robot when following optimal and neighboring trajectories. Energy consumption is reported for the motor side of the motor driver, when going from point $A$ to point $B$. Negative energy indicates energy being regenerated. The neighboring trajectories show a lower amount of total energy regeneration when compared to the optimal trajectory.}
\ra{1.3}
\begin{tabular}{@{} l c c c c @{}}\toprule
 && \multicolumn{3}{c}{$E_{A \to B} (J)$}\\
 \cmidrule{3-5}
 && Neighboring 1 & Neighboring 2 & Optimal\\
 \midrule
 Joint 1 &&  \underline{11.90}  &  12.03 & 12.04  \\
 Joint 2 &&  -25.19 & -25.83 & \underline{-28.18} \\
 Joint 3 &&  -0.66  & -1.19  & \underline{-1.31}  \\
 Total   &&  -13.95 & -14.99 & \underline{-17.45}  \\
 \bottomrule
\end{tabular} 
\label{table3}
\end{table}

\section{Conclusion}\label{sec:conclusion}
We investigated robotic systems having ultracapacitor based energy regenerative drive systems. For this purpose, a previously introduced framework is used to model the robot with the regenerative drive systems. Based on this model an optimization problem is formulated to find point-to-point trajectories maximizing energy regeneration. The PUMA 560 robot is used as a case study. The problem is solved numerically and the optimal trajectories are implemented on the PUMA 560 using a robust passivity based controller. Power flows are reported for the motor side and capacitor side of the motor driver. Experimental results show a good agreement with the theoretical results for the motor side of the motor driver and less agreement with the capacitor side. This is due to the efficiency of the motor driver and the power required to operate it. The motor drivers have a range for efficient operation, where operating outside of this range prevents energy from being regenerated back into the ultracapacitor. This is specially the case for Joint~3 of the robot where all the regenerated energy is dissipated in the motor driver. Also, while conducting the experiments, it was observed that controller chattering has a negative effect on energy regeneration. This could be due to the bandwidth limit for the motor drivers. In certain cases it might be necessary to compromise trajectory tracking for more energy regeneration. Using a higher quality motor driver with a higher bandwidth can also mitigate the problems associated with motor driver. On the other hand, including the inefficiencies of the motor driver in the model could provide more energy regeneration by prompting the optimization to look for different trajectories that operate in the efficient range for the motor drivers. Moreover, Experimental results for the neighboring trajectories showed the strong dependency of energy regeneration on trajectories followed by the robot joints thus, the need for trajectory optimization. Results also showed that a great portion of the energy is dissipated as mechanical losses due to the robots design. Even with these losses, energy regeneration resulted in about $13\%$ reduction in the overall energy consumption. In a factory assembly line with many robots, energy regeneration  can lead to significant reduction in operating cost. As part of future research paths, an alternative approach to the one taken here, could be to use model predictive control methods to provide optimal feedback directly, as opposed to solving for the optimal trajectory separately and enforcing it via a robust control method. Such an approach eliminates the need to reoptimize with changes in initial or final positions. In addition, we only consider  the star configuration for the semi-active joints. Using different configurations may lead to better results. This remains to be investigated in future work.

\appendices
\section{PUMA 560 Robot Model}\label{appen}
The $3 \times 13$ regressor matrix for the three main joints of the PUMA 560 robot (excluding the robot wrist) is given below where $Y_{ij}$ is the $i$-th row and $j$-th column element of the regressor matrix, $c_i=\cos(q_i)$, $s_i=\sin{q_i}$, $c_{ij}=\cos(q_i+q_j)$, and $s_{ij}=\sin(q_+q_j)$.
\begin{align*}
&Y_{11}=\ddot{q}_1&\\
&Y_{12}=\ddot{q}_1c_{2}^2 - 2\dot{q}_1\dot{q}_2c_{2}s_{2}\\
&Y_{13}=\ddot{q}_1c_{23}^2 - 2\dot{q}_1(\dot{q}_2+\dot{q}_3)s_{23}c_{23}\\
&Y_{14}=2(\ddot{q}_1- \dot{q}_1\dot{q}_2)c_{23}c_{2}  -2(\dot{q}_1\dot{q}_2+\dot{q}_1\dot{q}_3)s_{23}c_{2}\\
&Y_{15}=(\dot{q}_2+\dot{q}_3)^2c_{23} + (\ddot{q}_2 + \ddot{q}_3)s_{23}\\
&Y_{16}=Y_{17}=Y_{19}=Y_{110}=Y_{112}=Y_{113}=0\\
&Y_{18}=\dot{q}_2^2c_{2} + \ddot{q}_2s_{2}\\
&Y_{111}=\dot{q}_1\\
&Y_{22}=\dot{q}_1^2c_{2}s_{2}\\
&Y_{23}=\dot{q}_1^2c_{23}s_{23}\\
&Y_{21}=Y_{211}=Y_{213}=0\\
&Y_{24}=(c_{23}s_{2} + s_{23}c_{2})\dot{q}_1^2 - (2\dot{q}_3\dot{q}_2 + \dot{q}_3^2)s_{3}&\\
      &\qquad+(2\ddot{q}_2+\ddot{q}_3)c_{3}\\
&Y_{25}=\ddot{q}_1s_{23}\\
&Y_{26}=\ddot{q}_2\\
&Y_{27}=\ddot{q}_3\\
&Y_{28}=\ddot{q}_1s_{2}\\
&Y_{29}=-c_{23}\\
&Y_{210}=-c_{2}\\
&Y_{212}=\dot{q}_2\\
&Y_{31}=Y_{32}=Y_{36}=Y_{38}=Y_{310}=Y_{311}=Y_{312}=0\\
&Y_{33}=\dot{q}_1^2s_{23}c_{23}\\
&Y_{34}=\dot{q}_1^2s_{23}c_{2} + \dot{q}_2^2s_{3} + \ddot{q}_2c_{3}\\
&Y_{35}=\ddot{q}_1s_{23}\\
&Y_{37}=\ddot{q}_2+\ddot{q}_3\\
&Y_{39}=-c_{23}\\
&Y_{313}=\dot{q}_3\\
\end{align*}
The $13\times 1$ parameter vector, $\theta$, is given below where $m_i$ is the mass of the $i$-th robot link, $I_{ij}$ is the moment of inertia of the $i$-th link with respect to the $j$ axis of the a coordinate frame located at the center of mass of link $i$ and parallel to frame $i$, $\mathcal{C}_{ij}$ is the distance from the center of mass of link $i$, along the $j$ axis of frame $i$, to the origin of frame $i$, and $g$ is the gravity constant. Refer to Fig.~\ref{semiJMbond} and Fig.~\ref{puma1} for the definitions of the coordinate frames and other parameters. Numerical values used for the parameter vector are also given. 
\begin{align*}
&\theta_{1}=m_2d_2^2 + m_3(d_3 - d_2)^2 + I_{1y} + I_{2x} + I_{3x}\\
&\quad =2.8861\ \mathrm{kgm^2}\\
&\theta_{2}=m_2(\mathcal{C}_{2x}+A_2)^2+ m_3A_2^2 - I_{2x} + I_{2y}\\
&\quad=1.4425\ \mathrm{kgm^2}\\
&\theta_{3}=m_3\mathcal{C}_{3x}^2 - I_{3x} + I_{3y}=0.1990\ \mathrm{kgm^2}\\
&\theta_{4}=\mathcal{C}_{3x}A_2m_3=0.3815\ \mathrm{kgm^2}\\
&\theta_{5}=\mathcal{C}_{3x}m_3(d_3-d_2)=-0.1326\ \mathrm{kgm^2}\\
&\theta_{6}=m_2(\mathcal{C}_{2x}+A_2)^2 + m_3(\mathcal{C}_{3x}^2 +A_2^2) + I_{2z} + I_{3z}\\
&\quad=4.5860\ \mathrm{kgm^2}\\
&\theta_{7}=I_{3z}+m_3\mathcal{C}_{3x}^2=0.5945\ \mathrm{kgm^2}\\
&\theta_{8}=-d_2m_2(\mathcal{C}_{2x} + A_2) + A_2m_3(d_3 - d_2)\\
&\quad=-0.7938\ \mathrm{kgm^2}\\
&\theta_{9}=\mathcal{C}_{3x}gm_3=8.6677\ \mathrm{Nm}\\
&\theta_{10}=gm_2(\mathcal{C}_{2x} + A_2) + A_2gm_3=44.2165\ \mathrm{Nm}\\
&\theta_{11}=b_1+\frac{a_1^2}{R_1}=78.5975\ \mathrm{Nsm^{-1}}\\
&\theta_{12}=b_2+\frac{a_2^2}{R_2}=183.2162\ \mathrm{Nms}\\
&\theta_{13}=b_3+\frac{a_3^2}{R_3}=56.7933\ \mathrm{Nms}\\
\end{align*}
Link length values, $A_2$, $d_2$, and $d_3$ were taken from \cite{corke1994search} and verified by measuring the robot. All other parameters for the PUMA robot and the semi-active drive mechanisms were found by minimizing the difference between the measured robot outputs (e.g joint positions and motor driver voltages) and the robot model outputs. 

\section*{Acknowledgment}
This work was supported by the National Science Foundation, grants $\#1344954$ and $\#1536035$. We would also like to thank Rahul Harsha for his help in conducting the experiments and Farbod Rohani for his support in coding.

\bibliographystyle{ieeetr}
\bibliography{Bibliography_Dec-2017}

\begin{thebibliography}{10}

\bibitem{lukic2006power}
S.~M. Lukic, S.~G. Wirasingha, F.~Rodriguez, J.~Cao, and A.~Emadi, ``Power
  management of an ultracapacitor/battery hybrid energy storage system in an
  hev,'' in {\em 2006 IEEE Vehicle Power and Propulsion Conference}, pp.~1--6,
  IEEE, 2006.

\bibitem{hitt2009robotic}
J.~Hitt, T.~Sugar, M.~Holgate, R.~Bellman, and K.~Hollander, ``Robotic
  transtibial prosthesis with biomechanical energy regeneration,'' {\em
  Industrial Robot: An International Journal}, vol.~36, no.~5, pp.~441--447,
  2009.

\bibitem{shimizu2013super}
T.~Shimizu and C.~Underwood, ``Super-capacitor energy storage for
  micro-satellites: Feasibility and potential mission applications,'' {\em Acta
  Astronautica}, vol.~85, pp.~138--154, 2013.

\bibitem{richter2015framework}
H.~Richter, ``A framework for control of robots with energy regeneration,''
  {\em Journal of Dynamic Systems, Measurement, and Control}, vol.~137, no.~9,
  p.~091004, 2015.

\bibitem{conway2013electrochemical}
B.~E. Conway, {\em Electrochemical supercapacitors: scientific fundamentals and
  technological applications}.
\newblock Springer Science \& Business Media, 2013.

\bibitem{khaligh2010battery}
A.~Khaligh and Z.~Li, ``Battery, ultracapacitor, fuel cell, and hybrid energy
  storage systems for electric, hybrid electric, fuel cell, and plug-in hybrid
  electric vehicles: State of the art,'' {\em IEEE transactions on Vehicular
  Technology}, vol.~59, no.~6, pp.~2806--2814, 2010.

\bibitem{khoshnoud2015energy}
F.~Khoshnoud, Y.~Zhang, R.~Shimura, A.~Shahba, G.~Jin, G.~Pissanidis, Y.~K.
  Chen, and C.~W. De~Silva, ``Energy regeneration from suspension dynamic modes
  and self-powered actuation,'' {\em IEEE/ASME Transactions on Mechatronics},
  vol.~20, no.~5, pp.~2513--2524, 2015.

\bibitem{vinot2013optimal}
E.~Vinot and R.~Trigui, ``Optimal energy management of hevs with hybrid storage
  system,'' {\em Energy Conversion and Management}, vol.~76, pp.~437--452,
  2013.

\bibitem{song2014energy}
Z.~Song, H.~Hofmann, J.~Li, J.~Hou, X.~Han, and M.~Ouyang, ``Energy management
  strategies comparison for electric vehicles with hybrid energy storage
  system,'' {\em Applied Energy}, vol.~134, pp.~321--331, 2014.

\bibitem{rufer2002supercapacitor}
A.~Rufer and P.~Barrade, ``A supercapacitor-based energy-storage system for
  elevators with soft commutated interface,'' {\em IEEE Transactions on
  industry applications}, vol.~38, no.~5, pp.~1151--1159, 2002.

\bibitem{zhang2016high}
Z.~Zhang, X.~Zhang, W.~Chen, Y.~Rasim, W.~Salman, H.~Pan, Y.~Yuan, and C.~Wang,
  ``A high-efficiency energy regenerative shock absorber using supercapacitors
  for renewable energy applications in range extended electric vehicle,'' {\em
  Applied Energy}, vol.~178, pp.~177--188, 2016.

\bibitem{grbovic2011modeling}
P.~J. Grbovic, P.~Delarue, P.~Le~Moigne, and P.~Bartholomeus, ``Modeling and
  control of the ultracapacitor-based regenerative controlled electric
  drives,'' {\em IEEE Transactions on Industrial Electronics}, vol.~58, no.~8,
  pp.~3471--3484, 2011.

\bibitem{grbovic2011ultracapacitor}
P.~J. Grbovic, P.~Delarue, P.~Le~Moigne, and P.~Bartholomeus, ``The
  ultracapacitor-based controlled electric drives with braking and ride-through
  capability: Overview and analysis,'' {\em IEEE Transactions on Industrial
  Electronics}, vol.~58, no.~3, pp.~925--936, 2011.

\bibitem{kammer2016enhancing}
A.~S. Kammer and N.~Olgac, ``Enhancing energy harvesting capacity using delayed
  feedback control,'' in {\em 2016 American Control Conference (ACC)},
  pp.~1863--1868, IEEE, 2016.

\bibitem{asai2016nonlinear}
T.~Asai and J.~Scruggs, ``Nonlinear stochastic control of self-powered
  variable-damping vibration control systems,'' in {\em 2016 American Control
  Conference (ACC)}, pp.~442--448, IEEE, 2016.

\bibitem{carabin17}
G.~Carabin, E.~Wehrle, and R.~Vidoni, ``A review on energy-saving optimization
  methods for robotic and automatic systems,'' {\em Robotics}, vol.~6, no.~39,
  2017.

\bibitem{izumi1995optimal}
T.~Izumi, P.~Boyagoda, M.~Nakaoka, and E.~Hiraki, ``Optimal control of a servo
  system regenerating conservative energy to a condenser,'' in {\em Industrial
  Automation and Control: Emerging Technologies, 1995., International IEEE/IAS
  Conference on}, pp.~651--656, IEEE, 1995.

\bibitem{izumi2000energy}
T.~Izumi, ``Energy saving manipulator by regenerating conservative energy,'' in
  {\em Advanced Motion Control, 2000. Proceedings. 6th International Workshop
  on}, pp.~630--635, IEEE, 2000.

\bibitem{fujimoto2004minimum}
Y.~Fujimoto, ``Minimum energy biped running gait and development of energy
  regeneration leg,'' in {\em Advanced Motion Control, 2004. AMC'04. The 8th
  IEEE International Workshop on}, pp.~415--420, IEEE, 2004.

\bibitem{hansen2012enhanced}
C.~Hansen, J.~{\"O}ltjen, D.~Meike, and T.~Ortmaier, ``Enhanced approach for
  energy-efficient trajectory generation of industrial robots,'' in {\em
  Automation Science and Engineering (CASE), 2012 IEEE International Conference
  on}, pp.~1--7, IEEE, 2012.

\bibitem{hunter1981design}
B.~L. Hunter, {\em Design of a self-contained, active, regenerative computer
  controlled above-knee prosthesis}.
\newblock PhD thesis, Massachusetts Institute of Technology, 1981.

\bibitem{seth1987energy}
B.~Seth, {\em Energy regeneration and its application to active above-knee
  prostheses}.
\newblock PhD thesis, Massachusetts Institute of Technology, 1987.

\bibitem{tabor1988real}
K.~A. Tabor, {\em The real-time digital control of a regenerative above-knee
  prosthesis}.
\newblock PhD thesis, Massachusetts Institute of Technology, 1988.

\bibitem{hitt2010active}
J.~K. Hitt, T.~G. Sugar, M.~Holgate, and R.~Bellman, ``An active foot-ankle
  prosthesis with biomechanical energy regeneration,'' {\em Journal of medical
  devices}, vol.~4, no.~1, p.~011003, 2010.

\bibitem{hitt2007sparky}
J.~K. Hitt, R.~Bellman, M.~Holgate, T.~G. Sugar, and K.~W. Hollander, ``The
  sparky (spring ankle with regenerative kinetics) project: Design and analysis
  of a robotic transtibial prosthesis with regenerative kinetics,'' in {\em
  ASME 2007 International Design Engineering Technical Conferences and
  Computers and Information in Engineering Conference}, pp.~1587--1596,
  American Society of Mechanical Engineers, 2007.

\bibitem{holgate2008sparky}
M.~A. Holgate, J.~K. Hitt, R.~D. Bellman, T.~G. Sugar, and K.~W. Hollander,
  ``The sparky (spring ankle with regenerative kinetics) project: Choosing a dc
  motor based actuation method,'' in {\em 2008 2nd IEEE RAS \& EMBS
  International Conference on Biomedical Robotics and Biomechatronics},
  pp.~163--168, IEEE, 2008.

\bibitem{everarts2012variable}
C.~Everarts, B.~Dehez, and R.~Ronsse, ``Variable stiffness actuator applied to
  an active ankle prosthesis: Principle, energy-efficiency, and control,'' in
  {\em 2012 IEEE/RSJ International Conference on Intelligent Robots and
  Systems}, pp.~323--328, IEEE, 2012.

\bibitem{tucker2010mechanical}
M.~R. Tucker and K.~B. Fite, ``Mechanical damping with electrical regeneration
  for a powered transfemoral prosthesis,'' in {\em 2010 IEEE/ASME International
  Conference on Advanced Intelligent Mechatronics}, pp.~13--18, IEEE, 2010.

\bibitem{warner2015optimal}
H.~E. Warner, ``Optimal design and control of a lower-limb prosthesis with
  energy regeneration,'' Master's thesis, Cleveland State University, 2015.

\bibitem{warner2016IEEE}
H.~Warner, D.~Simon, and H.~Richter, ``Design optimization and control of a
  crank-slider actuator for a lower-limb prosthesis with energy regeneration,''
  in {\em IEEE International Conference on Advanced Intelligent Mechatronics,
  Banff, Canada}, 2016.

\bibitem{rarick2014optimal}
R.~Rarick, H.~Richter, A.~van~den Bogert, D.~Simon, H.~Warner, and T.~Barto,
  ``Optimal design of a transfemoral prosthesis with energy storage and
  regeneration,'' in {\em 2014 American Control Conference}, pp.~4108--4113,
  IEEE, 2014.

\bibitem{rohani2017optimal}
F.~Rohani, H.~Richter, and A.~J. Van~den Bogert, ``Optimal design and control
  of an electromechanical transfemoral prosthesis with energy regeneration,''
  {\em PloS one}, vol.~12, no.~11, p.~e0188266, 2017.

\bibitem{khalaf2018global}
P.~Khalaf and H.~Richter, ``On global, closed-form solutions to parametric
  optimization problems for robots with energy regeneration,'' {\em Journal of
  Dynamic Systems, Measurement, and Control}, vol.~140, no.~3, p.~031003, 2018.

\bibitem{khalaf2016parametric}
P.~Khalaf and H.~Richter, ``Parametric optimization of stored energy in robots
  with regenerative drive systems,'' in {\em 2016 IEEE International Conference
  on Advanced Intelligent Mechatronics (AIM)}, pp.~1424--1429, IEEE, 2016.

\bibitem{von2013numerical}
O.~von Stryk, ``Numerical solution of optimal control problems by direct
  collocation,'' {\em Optimal Control: Calculus of Variations, Optimal Control
  Theory and Numerical Methods}, vol.~111, p.~129, 2013.

\bibitem{richter2014semiactive}
H.~Richter, D.~Simon, and A.~van~den Bogert, ``Semiactive virtual control
  method for robots with regenerative energy-storing joints,'' in {\em Proc.
  19th IFAC World Congress, Cape Town, South Africa}, 2014.

\bibitem{SHV}
M.~Spong, S.~Hutchinson, and M.~Vidyasagar, {\em Robot Modeling and Control}.
\newblock Wiley, 2006.

\bibitem{karnopp2012system}
D.~C. Karnopp, D.~L. Margolis, and R.~C. Rosenberg, {\em System Dynamics:
  Modeling, Simulation, and Control of Mechatronic Systems}.
\newblock John Wiley \& Sons, 2012.

\bibitem{grbovic2013ultra}
P.~J. Grbovic, {\em Ultra-capacitors in power conversion systems: analysis,
  modeling and design in theory and practice}.
\newblock John Wiley \& Sons, 2013.

\bibitem{buller2001modeling}
S.~Buller, E.~Karden, D.~Kok, and R.~De~Doncker, ``Modeling the dynamic
  behavior of supercapacitors using impedance spectroscopy,'' in {\em
  Conference Record of the 2001 IEEE Industry Applications Conference. 36th IAS
  Annual Meeting}, vol.~4, pp.~2500--2504, IEEE, 2001.

\bibitem{chiang2013dynamic}
C.-J. Chiang, J.-L. Yang, and W.-C. Cheng, ``Dynamic modeling of the electrical
  and thermal behavior of ultracapacitors,'' in {\em 10th IEEE International
  Conference on Control and Automation (ICCA)}, pp.~1839--1844, IEEE, 2013.

\bibitem{bertrand2010fractional}
N.~Bertrand, J.~Sabatier, O.~Briat, and J.-M. Vinassa, ``Fractional non-linear
  modelling of ultracapacitors,'' {\em Communications in Nonlinear Science and
  Numerical Simulation}, vol.~15, no.~5, pp.~1327--1337, 2010.

\bibitem{kirk2012optimal}
D.~E. Kirk, {\em Optimal control theory: an introduction}.
\newblock Courier Corporation, 2012.

\bibitem{van2011implicit}
A.~J. Van Den~Bogert, D.~Blana, and D.~Heinrich, ``Implicit methods for
  efficient musculoskeletal simulation and optimal control,'' {\em Procedia
  IUTAM}, vol.~2, pp.~297--316, 2011.

\bibitem{code}
{Control, Robotics and Mechatronics lab}, ``Code for energy regeneration
  optimization.'' \url{http://academic.csuohio.edu/richter_h/lab/regen/}, 2018.
\newblock accessed 23-February-2018.

\bibitem{wachter2006implementation}
A.~W{\"a}chter and L.~T. Biegler, ``On the implementation of an interior-point
  filter line-search algorithm for large-scale nonlinear programming,'' {\em
  Mathematical programming}, vol.~106, no.~1, pp.~25--57, 2006.

\bibitem{musolino2013new}
V.~Musolino, L.~Piegari, and E.~Tironi, ``New full-frequency-range
  supercapacitor model with easy identification procedure,'' {\em IEEE
  Transactions on Industrial Electronics}, vol.~60, no.~1, pp.~112--120, 2013.

\bibitem{corke1994search}
P.~I. Corke and B.~Armstrong-Helouvry, ``A search for consensus among model
  parameters reported for the puma 560 robot,'' in {\em IEEE International
  Conference on Robotics and Automation}, pp.~1608--1613, IEEE, 1994.

\end{thebibliography}

\end{document}